\documentstyle[12pt]{article}

\catcode`\@=11
\def\numberbysection{\@addtoreset{equation}{section}
        \def\theequation{\thesection.\arabic{equation}}}
\numberbysection

\begin{document}

\newlength{\lno} \lno1.5cm \newlength{\len} \len=\textwidth%
\addtolength{\len}{-\lno}

\baselineskip7mm \renewcommand{\thefootnote}{\fnsymbol{footnote}} \newpage %
\setcounter{page}{0} 
\begin{titlepage}     
\vspace{0.5cm}
\begin{center}
{\Large\bf Bethe Ans\"{a}tze for 19-vertex Models}\\
\vspace{1cm}
{\large A. Lima-Santos } \\
\vspace{1cm}
{\large \em Universidade Federal de S\~ao Carlos, Departamento de F\'{\i}sica \\
Caixa Postal 676, CEP 13569-905~~S\~ao Carlos, Brasil}\\
\end{center}
\vspace{1.2cm}
\begin{abstract}
The nineteen-vertex models of Zalomodchikov-Fateev, Izergin-Korepin and the
supersymmetric $osp(1|2)$ with periodic boundary conditions are studied. We
find the spectrum of these quantum spin chain using the Coordinate Bethe
Ansatz. The approach is a suitable parametrization of their wavefunctions. We
also applied the Algebraic Bethe Ansatz in order to obtain the eigenvalues 
and eigenvectors of the corresponding transfer matrices.
\end{abstract}
\vspace{1.2cm}
\begin{center}
\end{center}
\vfill
\centerline{\today}
\end{titlepage}

\renewcommand{\thefootnote}{\arabic{footnote}} \setcounter{footnote}{0}

\newpage{}

\section{Introduction}

One-dimensional quantum spin chain Hamiltonians and classical statistical
systems in two spatial dimensions on a lattice (vertex models), share a
common mathematical structure responsible by our understanding of these
integrable models \cite{Baxter, KIB}. If the Boltzmann weights underlying
the vertex models are obtained from solutions of the Yang-Baxter ({\small YB}%
) equation the commutativity of the associated transfer matrices immediately
follow, leading to their integrability.

The Bethe Ansatz ({\small BA}) is the powerful method in the analysis of
integrable quantum models. There are several versions: Coordinate {\small BA}
\cite{Bethe}, Algebraic{\small \ BA} \cite{FT}, Analytical {\small BA} \cite
{VR}, etc. developed for diagonalization of the corresponding Hamiltonian.

The simplest version is the Coordinate {\small BA.}\ In this framework one
can obtain the eigenfunctions and the spectrum of the Hamiltonian from its
eigenvalue problem. It is really simple and clear for the two-state models
like the six-vertex models but becomes awkward for models with a higher
number of states.

The Algebraic {\small BA}, also proverbial as Quantum Inverse Scattering
method, is an elegant and important generalization of the Coordinate {\small %
BA}. It is based on the idea of constructing eigenfunctions of the
Hamiltonian via creation and annihilation operators acting on a reference
state. Here one uses the miraculous fact that the {\small YB} equation can
be recast in the form of commutation relations for the matrix elements of
the monodromy matrix which play the role of creation and annihilation
operators. From this monodromy matrix we get the transfer matrix which, by
construction, commutes with the Hamiltonian. Thus, constructing
eigenfunctions of the transfer matrix determines the eigenfunctions of the
Hamiltonian.

Imposing appropriate boundary conditions the {\small BA} method leads to a
system of equations, the {\small BA} equations, which are useful in the
thermodynamic limit. The energy of the ground state and its excitations,
velocity of sound, etc., may be calculated in this limit. Moreover, in
recent years we witnessed another very fruitful connection between the 
{\small BA} method and conformal field theory. \ Using the Algebraic {\small %
BA}, Korepin \cite{KO} found various representations of correlators in
integrable models and more recently Babujian and Flume \cite{BF} developed a
method from the Algebraic {\small BA} which reveals a link to the Gaudin
model and render in the quasiclassical limit solutions of the
Knizhnik-Zamolodchikov equations for the $SU(2)$ Wess-Zumino-Novikov-Witten
conformal theory.

Integrable quantum systems containing Fermi fields have been attracting
increasing interest due to their potential applications in condensed matter
physics. The prototypical examples of such systems are the supersymmetric
generalizations of the Hubbard and $t$-$J$ models \cite{EK0}. They lead to a
generalization of the {\small YB} equation associated with the introduction
of the a $Z_{2}$ grading \cite{KS1} which leads to appearance of additional
signs in the {\small YB} equation.

In this paper we consider the Coordinate and Algebraic versions of the 
{\small BA} for the trigonometric three-state vertex models of $19$-vertices
with periodic boundary conditions. These models are well-known in the
literature: the Zamolodchikov-Fateev ({\small ZF}) model or $A_{1}^{1}$
model \cite{ZF}, the Izergin-Korepin ({\small IK}) model or $A_{2}^{2}$
model \cite{IK} and the supersymmetric $osp(1|2)$ model \cite{BS}.

While the {\small BA} solution of the periodic {\small ZF} model was derived
by a fusion procedure \cite{KRS} in \cite{BT} and \cite{KR}, a
generalization of the Algebraic {\small BA} was developed by Tarasov \cite
{TA} to solve the {\small IK} model. Moreover, the {\small IK} model was
solved via Coordinate BA by Batchelor {\it at al} in \cite{BAT1}.

In the context of the Algebraic {\small BA} , the version presented here is
based on the Tarosov approach but now we include the {\small ZF} model and
we also extend it to the graded version of the quantum inverse scattering
method in order to consider the $osp(1|2)$ model.

In the context of the Coordinate {\small BA}, we propose here a new
parametrization of wavefunctions. This result is of fundamental importance
since it allows us to treat these $19$-vertex models in the same way and the
Coordinate {\small BA} for these three-states models becomes simple enough
as for two-state models.

The main goal in this paper is to reveal the common structure of these $19$%
-vertex models which permits us to apply the {\small BA} method, unifying
old and new results.

The paper is organized as follows: In Section $2$ we present the models. In
Section $3$ the spectra of the corresponding Hamiltonians are derived using
the Coordinate {\small BA} and in Section $4$ the Algebraic {\small BA} is
also used to diagonalize the corresponding transfer matrices. We justify
this twofold presentation remarking that the {\small BA} method is
apparently version dependent. It means that when one solves a model using a
particular {\small BA} version is not always clear how to extend the
solution for all possible versions. For example, the biquadratic model was
solved by Coordinate {\small BA} in \cite{PAR, KLS} and its Algebraic 
{\small BA} version is still unknown. Finally, the conclusions are reserved
for Section $5$.

\section{Description of the models}

Let us start with the graded formulation and then recover the non graded
from it.

Consider $V=V_{0}\oplus V_{1}$ a $Z_{2}$-graded vector space where $0$ and $%
1 $ denote the even and odd parts respectively. The multiplication rules in
the graded tensor product space $V\stackrel{s}{\otimes }V$ differ from the
ordinary ones by the appearance of additional signs. The components of a
linear operator $A\stackrel{s}{\otimes }B\in V\stackrel{s}{\otimes }V$
result in matrix elements of the form 
\begin{equation}
(A\stackrel{s}{\otimes }B)_{\alpha \beta }^{\gamma \delta }=(-)^{p(\beta
)(p(\alpha )+p(\gamma ))}\ A_{\alpha \gamma }B_{\beta \delta }  \label{eq2.1}
\end{equation}
The action of the graded permutation operator ${\cal P}$ on the vector $%
\left| \alpha \right\rangle \stackrel{s}{\otimes }\left| \beta \right\rangle
\in V\stackrel{s}{\otimes }V$ is defined by 
\begin{equation}
{\cal P}\ \left| \alpha \right\rangle \stackrel{s}{\otimes }\left| \beta
\right\rangle =(-)^{p(\alpha )p(\beta )}\left| \beta \right\rangle \stackrel{%
s}{\otimes }\left| \alpha \right\rangle \Longrightarrow ({\cal P})_{\alpha
\beta }^{\gamma \delta }=(-)^{p(\alpha )p(\beta )}\delta _{\alpha \delta }\
\delta _{\beta \gamma }  \label{eq2.2}
\end{equation}
where $p(\alpha )=1\ (0)$ if $\left| \alpha \right\rangle $ is an odd (even)
element.

The central object in the theory of integrable models is the ${\cal R}$%
-matrix ${\cal R}(\lambda )$, where $\lambda $ is the spectral parameter. It
acts on the tensor product $V^{1}\otimes V^{2}$ for a given vector space $V$
and it is solution of the Yang-Baxter ({\small YB}) equation 
\begin{equation}
{\cal R}_{12}(\lambda ){\cal R}_{13}(\lambda +\mu ){\cal R}_{23}(\mu )={\cal %
R}_{23}(\mu ){\cal R}_{13}(\lambda +\mu ){\cal R}_{12}(\lambda )
\label{eq2.3}
\end{equation}
in $V^{1}\otimes V^{2}\otimes V^{3}$, where ${\cal R}_{12}={\cal R}\otimes 
{\cal I}$, ${\cal R}_{23}={\cal I}\otimes {\cal R}$, etc.

In the graded case, ${\cal R}_{13}$ however, does not act trivially on the
second space due to signs generated by commuting odd operators. In this
case, the graded {\small YB} equation in components reads 
\begin{eqnarray}
&&{\cal R}_{ii^{\prime }}^{kk^{\prime }}(\lambda ){\cal R}_{ki^{^{\prime
\prime }}}^{jk^{^{\prime \prime }}}(\lambda +\mu ){\cal R}_{k^{\prime
}k^{^{\prime \prime }}}^{j^{\prime }j^{^{\prime \prime }}}(\mu
)(-)^{p(k^{\prime })(p(i^{^{\prime \prime }})+p(k^{^{\prime \prime }}))} 
\nonumber \\
&=&{\cal R}_{i^{\prime }i^{^{\prime \prime }}}^{k^{\prime }k^{^{\prime
\prime }}}(\mu ){\cal R}_{ik^{^{\prime \prime }}}^{kj^{^{\prime \prime
}}}(\lambda +\mu ){\cal R}_{kk^{^{\prime }}}^{jj^{^{\prime }}}(\lambda
)(-)^{p(k^{^{\prime }})(p(j^{^{\prime \prime }})+p(k^{^{\prime \prime }}))}
\label{eq2.4}
\end{eqnarray}
Besides ${\cal R}$ we have to consider matrices $R={\cal PR}$ which satisfy 
\begin{equation}
R_{12}(\lambda )R_{23}(\lambda +\mu )R_{12}(\mu )=R_{23}(\mu )R_{12}(\lambda
+\mu )R_{23}(\lambda )  \label{eq2.5}
\end{equation}
Because only $R_{12}$ and $R_{23}$ are involved, Eq.(\ref{eq2.5}) written in
components looks the same as in the non graded case. Moreover, the matrices $%
{\cal R}_{ng}=PR$ satisfy the ordinary {\small YB} equation (\ref{eq2.3})
where $P$ is the non graded permutation operator.

\subsection{The $R$-matrices}

\bigskip We will consider 19-vertex models for which their $R$ matrices have
a common form 
\begin{equation}
R(\lambda )=\left( 
\begin{array}{lllllllllll}
x_{1} & 0 & \ 0 &  & 0 & \ 0 & 0 &  & \ 0 & 0 & 0 \\ 
0 & y_{5} & \ 0 &  & x_{2} & \ 0 & 0 &  & \ 0 & 0 & 0 \\ 
0 & 0 & \ y_{7} &  & 0 & \ y_{6} & 0 &  & \ x_{3} & 0 & 0 \\ 
&  &  &  &  &  &  &  &  &  &  \\ 
0 & x_{2} & \ 0 &  & x_{5} & \ 0 & 0 &  & 0 & 0 & 0 \\ 
0 & 0 & \varepsilon y_{6} &  & 0 & \varepsilon x_{4} & 0 &  & \varepsilon
x_{6} & 0 & 0 \\ 
0 & 0 & \ 0 &  & 0 & \ 0 & y_{5} &  & \ 0 & x_{2} & 0 \\ 
&  &  &  &  &  &  &  &  &  &  \\ 
0 & 0 & \ x_{3} &  & 0 & \ x_{6} & 0 &  & \ x_{7} & 0 & 0 \\ 
0 & 0 & \ 0 &  & 0 & \ 0 & x_{2} &  & \ 0 & x_{5} & 0 \\ 
0 & 0 & \ 0 &  & 0 & \ 0 & 0 &  & \ 0 & 0 & x_{1}
\end{array}
\right)  \label{eq2.6}
\end{equation}
Here we have assumed that the grading of threefold space is $p(1)=1$, $%
p(2)=\varepsilon$ and $p(3)=1$, where $\varepsilon=\pm $. The matrix
elements $x_{i}$ and $y_{i}$ for each model will be listed below .

\begin{itemize}
\item  {\em ZF }${\em R}${\em -matrix}
\end{itemize}

The simplest $19$-vertex model is the {\small ZF} model or $A_{1}^{1}$ model
. The solution of the {\small YB} equation was found in \cite{ZF}. It can
also be constructed from the six-vertex model using the fusion procedure.
The corresponding $R$-matrix has the form (\ref{eq2.6}) with $\varepsilon =1$
and 
\begin{eqnarray}
x_{1}(\lambda ) &=&\sinh (\lambda +\eta )\sinh (\lambda +2\eta ),\quad
x_{2}(\lambda )=\sinh \lambda \sinh (\lambda +\eta )  \nonumber \\
x_{3}(\lambda ) &=&\sinh \lambda \sinh (\lambda -\eta ),\quad x_{5}(\lambda
)=y_{5}(\lambda )=\sinh (\lambda +\eta )\sinh 2\eta  \nonumber \\
x_{6}(\lambda ) &=&y_{6}(\lambda )=\sinh \lambda \sinh 2\eta ,\quad
x_{7}(\lambda )=y_{7}(\lambda )=\sinh \eta \sinh 2\eta  \nonumber \\
x_{4}(\lambda ) &=&x_{2}(\lambda )+x_{7}(\lambda )  \label{RZF}
\end{eqnarray}

\begin{itemize}
\item  {\em IK \ }${\em R}${\em -matrix}
\end{itemize}

The solution of this {\small YB} equation was found in \cite{IK}. It can not
be constructed from the six-vertex model using the fusion procedure. The $R$%
-matrix has the form (\ref{eq2.6}) with $\varepsilon =1$ and 
\begin{eqnarray}
x_{1}(\lambda ) &=&\sinh (\lambda -5\eta )+\sinh \eta ,\quad x_{2}(\lambda
)=\sinh (\lambda -3\eta )+\sinh 3\eta  \nonumber \\
x_{3}(\lambda ) &=&\sinh (\lambda -\eta )+\sinh \eta  \nonumber \\
x_{4}(\lambda ) &=&\sinh (\lambda -3\eta )+\sinh 3\eta -\sinh 5\eta +\sinh
\eta  \nonumber \\
x_{5}(\lambda ) &=&-\sinh 2\eta ({\rm e}^{-\lambda +3\eta }+{\rm e}^{-3\eta
}),\quad y_{5}(\lambda )=-\sinh 2\eta ({\rm e}^{\lambda -3\eta }+{\rm e}%
^{3\eta })  \nonumber \\
x_{6}(\lambda ) &=&{\rm e}^{2\eta }\sinh 2\eta (1-{\rm e}^{-\lambda }),\quad
y_{6}(\lambda )={\rm e}^{-2\eta }\sinh 2\eta (1-{\rm e}^{\lambda }) 
\nonumber \\
x_{7}(\lambda ) &=&-2{\rm e}^{-\lambda +2\eta }\sinh \eta \sinh 2\eta -{\rm e%
}^{-\eta }\sinh 4\eta  \nonumber \\
y_{7}(\lambda ) &=&2{\rm e}^{\lambda -2\eta }\sinh \eta \sinh 2\eta -{\rm e}%
^{\eta }\sinh 4\eta  \label{RIK}
\end{eqnarray}

\begin{itemize}
\item  {\em \ }$Osp(1|2)${\em \ }${\em R}${\em -matrix}
\end{itemize}

The trigonometric solution of the graded {\small YB} equation for the
fundamental representation of $osp(1|2)$ algebra was found by Bazhanov and
Shadrikov in \cite{BS}. It has the form (\ref{eq2.6}) with \ $\varepsilon
=-1 $ and 
\begin{eqnarray}
x_{1}(\lambda ) &=&\sinh (\lambda +2\eta )\sinh (\lambda +3\eta ),\quad
x_{2}(\lambda )=\sinh \lambda \sinh (\lambda +3\eta )  \nonumber \\
x_{3}(\lambda ) &=&\sinh \lambda \sinh (\lambda +\eta )  \nonumber \\
x_{4}(\lambda ) &=&\sinh \lambda \sinh (\lambda +3\eta )-\sinh 2\eta \sinh
3\eta  \nonumber \\
x_{5}(\lambda ) &=&{\rm e}^{-\lambda /3}\sinh 2\eta \sinh (\lambda +3\eta
),\quad y_{5}(\lambda )={\rm e}^{\lambda /3}\sinh 2\eta \sinh (\lambda
+3\eta )  \nonumber \\
x_{6}(\lambda ) &=&-{\rm e}^{-\lambda /3-2\eta }\sinh 2\eta \sinh \lambda
,\quad y_{6}(\lambda )={\rm e}^{\lambda /3+2\eta }\sinh 2\eta \sinh \lambda 
\nonumber \\
x_{7}(\lambda ) &=&{\rm e}^{\lambda /3}\sinh 2\eta \left( \sinh (\lambda
+3\eta )+{\rm e}^{-\eta }\sinh \lambda \right)  \nonumber \\
y_{7}(\lambda ) &=&{\rm e}^{-\lambda /3}\sinh 2\eta \left( \sinh (\lambda
+3\eta )+{\rm e}^{\eta }\sinh \lambda \right)  \label{ROSP}
\end{eqnarray}

The rational limit of (\ref{ROSP}) is well-known in the literature \cite{Ku2}
and can be written in the form : 
\begin{equation}
R(\lambda ,\eta )=\eta {\cal I}+\lambda {\cal P}+\frac{\lambda \eta }{%
\lambda +\frac{3\eta }{2}}{\cal U}  \label{eq2.8}
\end{equation}
where ${\cal I}$ is the identity operator, ${\cal P}$ is the graded
permutation operator (\ref{eq2.2}) and ${\cal U}$ is the rank-one projector $%
{\cal U}^{2}={\cal U}$. \ The algebraic solution of (\ref{eq2.8}) was
obtained by Martins \cite{MA}, as a limit of the algebraic solution of the 
{\small IK} model.

\subsection{The Hamiltonians}

In order to derive the Hamiltonian, it is convenient to expand the $R$%
-matrix around the regular point $\lambda =0$. For the $19$-vertex models
the corresponding solutions with the standard normalization can be read
directly from (\ref{eq2.6}). They have the form 
\begin{equation}
R(\lambda ,\eta )\sim {\cal I}+\lambda (\alpha ^{-1}{\cal H}+\beta {\cal I})+%
{\rm o}(\lambda ^{2}).  \label{eq2.9}
\end{equation}
with $\alpha $ and $\beta $ being scalar functions.

The Hamiltonian is a local sum given by 
\begin{equation}
H=\sum_{k=1}^{N-1}H_{k,k+1}+H_{N,1}  \label{eq2.10}
\end{equation}
where $H_{k,k+1}$ is the ${\cal H}$ in (\ref{eq2.9}) acting on the quantum
spaces at sites $k$ and $k+1$. Using a spin language, this is a spin $1$
Hamiltonian. In the basis where $S_{k}^{z}$ is diagonal with eigenvectors $%
\left| +,k\right\rangle ,\left| 0,k\right\rangle ,\left| -,k\right\rangle $
and eigenvalues $1,0,-1$, respectively, the Hamiltonian densities acting on
two neighboring sites are then given by 
\begin{equation}
H_{k,k+1}= 
\begin{array}{l}
\left| ++\right\rangle \\ 
\left| +0\right\rangle \\ 
\left| +{\scriptsize \ }-\right\rangle \\ 
\left| 0+\right\rangle \\ 
\left| 0{\scriptsize \ }0\right\rangle \\ 
\left| 0-\right\rangle \\ 
\left| -+\right\rangle \\ 
\left| -0\right\rangle \\ 
\left| --\right\rangle
\end{array}
\left( 
\begin{array}{lllllllll}
z_{1} & 0 & \ 0 & 0 & \ 0 & 0 & \ 0 & 0 & 0 \\ 
0 & \stackrel{\_}{z}_{5} & \ 0 & 1 & \ 0 & 0 & \ 0 & 0 & 0 \\ 
0 & 0 & \ \stackrel{\_}{z}_{7} & 0 & \ \stackrel{\_}{z}_{6} & 0 & \ z_{3} & 0
& 0 \\ 
0 & 1 & \ 0 & z_{5} & \ 0 & 0 & 0 & 0 & 0 \\ 
0 & 0 & \varepsilon \stackrel{\_}{z}_{6} & 0 & \varepsilon z_{4} & 0 & 
\varepsilon z_{6} & 0 & 0 \\ 
0 & 0 & \ 0 & 0 & \ 0 & \stackrel{\_}{z}_{5} & \ 0 & 1 & 0 \\ 
0 & 0 & \ z_{3} & 0 & \ z_{6} & 0 & \ z_{7} & 0 & 0 \\ 
0 & 0 & \ 0 & 0 & \ 0 & 1 & \ 0 & z_{5} & 0 \\ 
0 & 0 & \ 0 & 0 & \ 0 & 0 & \ 0 & 0 & z_{1}
\end{array}
\right) _{k,k+1}  \label{eq2.11}
\end{equation}
where the matrix elements for each model are:

\begin{itemize}
\item  {\em ZF Hamiltonian}
\end{itemize}

\ For the {\small ZF} model the corresponding quantum spin chain is the spin 
$1$ {\small XXZ} model. The two site Hamiltonian is derived from (\ref{eq2.9}%
) and has the form (\ref{eq2.11}) with 
\begin{eqnarray}
\varepsilon &=&1,\quad \alpha =\sinh 2\eta ,\quad \beta =0  \nonumber \\
z_{1} &=&0,\quad z_{3}=-1,\quad z_{4}=-2\cosh 2\eta ,  \nonumber \\
z_{5} &=&\stackrel{\_}{z}_{5}=-\cosh 2\eta ,\quad z_{6}=\stackrel{\_}{z}%
_{6}=2\cosh \eta  \nonumber \\
z_{7} &=&\stackrel{\_}{z}_{7}=-1-2\cosh 2\eta  \label{HZF}
\end{eqnarray}

\begin{itemize}
\item  {\em IK Hamiltonian}
\end{itemize}

In the {\small IK} model the two-site Hamiltonian for the corresponding
quantum chain has the form (\ref{eq2.11}) with 
\begin{eqnarray}
\varepsilon &=&1,\quad \alpha =-2\sinh 2\eta ,\quad \beta =0  \nonumber \\
z_{1} &=&0,\quad z_{3}=\frac{\cosh \eta }{\cosh 3\eta },\quad z_{4}=-2\frac{%
\sinh 4\eta \sinh \eta }{\cosh 3\eta },  \nonumber \\
z_{5} &=&-{\rm e}^{-2\eta },\quad \stackrel{\_}{z}_{5}=-{\rm e}^{2\eta } 
\nonumber \\
z_{6} &=&{\rm e}^{2\eta }\frac{\sinh 2\eta }{\cosh 3\eta },\qquad \stackrel{%
\_}{z}_{6}=-{\rm e}^{-2\eta }\frac{\sinh 2\eta }{\cosh 3\eta }  \nonumber \\
z_{7} &=&-{\rm e}^{-4\eta }\frac{\cosh \eta }{\cosh 3\eta },\quad \stackrel{%
\_}{z}_{7}=-{\rm e}^{4\eta }\frac{\cosh \eta }{\cosh 3\eta }  \label{HIK}
\end{eqnarray}

\begin{itemize}
\item  $Osp(1|2)${\em \ Hamiltonian}
\end{itemize}

The two site quantum Hamiltonian associated with the $osp(1|2)$ model has
the form (\ref{eq2.11}) with

\begin{eqnarray}
\varepsilon &=&-1,\quad \alpha =\sinh 2\eta ,\quad \beta =-\coth 2\eta \qquad
\nonumber \\
z_{1} &=&\cosh 2\eta ,\quad z_{3}=\frac{\sinh \eta }{\sinh 3\eta },\quad
z_{4}=1+\coth 3\eta \sinh 2\eta ,  \nonumber \\
z_{5} &=&-\frac{\sinh 2\eta }{3},\quad \stackrel{\_}{z}_{5}=-z_{5}  \nonumber
\\
z_{6} &=&-{\rm e}^{-2\eta }\frac{\sinh 2\eta }{\sinh 3\eta },\qquad 
\stackrel{\_}{z}_{6}={\rm e}^{2\eta }\frac{\sinh 2\eta }{\sinh 3\eta } 
\nonumber \\
z_{7} &=&\frac{\sinh 2\eta }{3}+{\rm e}^{-\eta }\frac{\sinh 2\eta }{\sinh
3\eta },\quad \stackrel{\_}{z}_{7}=-\frac{\sinh 2\eta }{3}+{\rm e}^{\eta }%
\frac{\sinh 2\eta }{\sinh 3\eta }  \label{HOSP}
\end{eqnarray}

Having now built a common ground for these\ models, we may proceed to find
their spectra. We begin with the Coordinate {\small BA} because of its
simplicity.

\section{The Coordinate Bethe Ansatz}

In this section results are presented for a periodic quantum spin chain of $%
N $ atoms each with spin $1$ described by the Hamiltonian (\ref{eq2.10}). At
each site, the {\em spin variable} may be $+1,0,-1$, so that the Hilbert
space of the spin chain is ${\cal H}^{(N)}=\otimes ^{N}V$ where $V=C^{3}$
with basis $\left\{ \left| +\right\rangle ,\left| 0\right\rangle ,\left|
-\right\rangle \right\} $. The dimension of the Hilbert space is {\rm dim}$%
{\cal H}^{(N)}=3^{N}$. On ${\cal H}^{(N)}$ we consider the Hamiltonians
presented in the previous section.

From (\ref{eq2.10}) one can see that $H$ commutes with the operator which
shifts the states of the chain by one unity. It means translational
invariance of $H$. Moreover, the Hamiltonian (\ref{eq2.10}) preserves the
third component of the {\em spin} 
\begin{equation}
\left[ H,S_{T}^{z}\right] =0,\qquad S_{T}^{z}=\sum_{k=1}^{N}S_{k}^{z}
\label{eq3.1}
\end{equation}
This allows us to divide the Hilbert space of states into different sectors,
each labelled by the eigenvalue of the operator number $r=N-S_{T}^{z}$ . We
shall denote by ${\cal H}_{n}^{(N)}$ the subspace of ${\cal H}^{(N)}$ with $%
r=n$. One can easily see that ${\rm dim}{\cal H}^{(N)}=\sum_{r=0}^{N}{\rm dim%
}{\cal H}_{r}^{(N)}$ with 
\begin{equation}
{\rm dim}{\cal H}_{r}^{(N)}=\sum_{k=0}^{[\frac{r}{2}]}\left( 
\begin{array}{c}
N \\ 
r-2k
\end{array}
\right) \left( 
\begin{array}{c}
N-r+2k \\ 
k
\end{array}
\right) ,
\end{equation}
where $[\frac{r}{2}]$ means the integer part of $\frac{r}{2}$.

\subsection{Sector r=0}

The sector ${\cal H}_{0}^{(N)}$ contains only one state, the {\em reference
state}, with all spin value equal to $+1$, $\Psi _{0}=\prod_{k}\left|
+,k\right\rangle $, satisfying $H\Psi _{0}=E_{0}\Psi _{0}$, with $%
E_{0}=Nz_{1}$. All other energies will be measured relative to this state.
It means that we will seek eigenstates of $H$ satisfying $(H-Nz_{1})\Psi
_{r}=\epsilon _{r}\Psi _{r}$ , in every sector $r$.

\subsection{Sector r=1}

In ${\cal H}_{1}^{(N)}${\it ,} the subspace of states with all spin value
equal to $+1$ except one with value $0$. There are $N$ states $\left|
k[0]\right\rangle =\left| +++\,{%
\raisebox{-0.65em}{$\stackrel{\textstyle
0}{\scriptstyle k}$}}\,++\cdots +\right\rangle $ which span a basis of $%
{\cal H}_{1}^{(N)}$. The ansatz for the eigenstate is thus of the form 
\begin{equation}
\Psi _{1}=\sum_{k=1}^{N}A(k)\left| k[0]\right\rangle  \label{eq3.2}
\end{equation}
The unknown wavefunction $A(k)$ determines the probability that the {\em %
spin variable} has the value $0$ at the site $k$.

From the complete invariance translational due to the periodic boundary
conditions, it follows that $A(k)$ is just the wavefunction for a plane wave 
\begin{equation}
A(k)=\xi ^{k}  \label{eq3.3}
\end{equation}
where $\xi ={\rm e}^{i\theta }$ , $\theta $ being some particular momentum
fixed by the boundary condition $A(N+k)=A(k)$.

When $H$ acts on $\left| k[0]\right\rangle $ , it sees the reference
configuration, except in the vicinity of $k$, and using (\ref{eq2.11}) we
obtain the eigenvalue equations 
\begin{equation}
(\epsilon _{1}+2z_{1}-z_{5}-\stackrel{\_}{z}_{5})A(k)=A(k-1)+A(k+1)
\label{eq3.4}
\end{equation}
The plane wave parametrization (\ref{eq3.3}) solves (\ref{eq3.4}) provided 
\begin{equation}
\epsilon _{1}=-2z_{1}+z_{5}+\stackrel{\_}{z}_{5}+\xi +\xi ^{-1}
\label{eq3.5}
\end{equation}
Thus $\Psi _{1}$ is the eigenstate of $H$ in the sector $r=1$ with
eigenvalue $E_{1}=(N-2z_{1})+z_{5}+\stackrel{\_}{z}_{5}+2\cos \theta $,
where $\theta =2\pi l/N$ , $l=0,1,...,N-1$.

\subsection{Sector r=2}

In the Hilbert space ${\cal H}_{2}^{(N)}$ we have $N$ states of the type $%
\left| k[-]\right\rangle =\left| ++\,{\raisebox{-0.65em}{$\stackrel{%
\textstyle -}{\scriptstyle k}$}}\,++\cdots +\right\rangle $ and $N(N-1)/2$
states of the type $\left| k_{1}[0],k_{2}[0]\right\rangle =\left| ++\,{%
\raisebox{-0.65em}{$\stackrel{\textstyle 0}{\scriptstyle k_1}$}}\,++\,{%
\raisebox{-0.65em}{$\stackrel{\textstyle 0}{\scriptstyle k_2}$}}\,++\cdots
+\right\rangle $. We seek these eigenstates in the form 
\begin{equation}
\Psi _{2}=\sum_{k_{1}<k_{2}}A(k_{1},k_{2})\left|
k_{1}[0],k_{2}[0]\right\rangle +\sum_{k=1}^{N}B(k)\left| k[-]\right\rangle .
\label{eq3.6}
\end{equation}

The periodicity condition reads now 
\begin{equation}
A(k_{2},N+k_{1})=A(k_{1},k_{2})\qquad {\rm and}\qquad B(N+k)=B(k)
\label{eq3.7}
\end{equation}
Following Bethe \cite{Bethe}, the wavefunction $A(k_{1},k_{2})$ can be
parametrized using the superposition of plane waves (\ref{eq3.3}) including
the scattering of two {\em pseudoparticles} with {\em momenta} $\theta _{1}$
and $\theta _{2}$, ($\xi _{j}=e^{i\theta _{j}},j=1,2$): 
\begin{equation}
A(k_{1},k_{2})=A_{12}\xi _{1}^{k_{1}}\xi _{2}^{k_{2}}+A_{21}\xi
_{1}^{k_{2}}\xi _{2}^{k_{1}}  \label{eq3.8}
\end{equation}
which satisfy the periodic boundary condition (\ref{eq3.7}) provided 
\begin{equation}
A_{12}=A_{21}\xi _{1}^{N}\quad ,\quad A_{21}=A_{12}\xi _{2}^{N}
\label{eq3.9}
\end{equation}
and the parametrization of $B(k)$ is still undetermined at this stage.

Before we try to parametrize $B(k)$ let us consider the Schr\"{o}dinger
equation $(E_{2}-Nz_{1})\Psi _{2}=\epsilon _{2}\Psi _{2}$ . From the
explicit form of $H$ acting on two sites (\ref{eq2.11}) we derive the
following set of eigenvalue equations 
\begin{eqnarray}
(\epsilon _{2}+4z_{1}-2z_{5}-2\stackrel{\_}{z}_{5})A(k_{1},k_{2})
&=&A(k_{1}-1,k_{2})+A(k_{1}+1,k_{2})  \nonumber \\
&&+A(k_{1},k_{2}-1)+A(k_{1},k_{2}+1)  \label{eq3.10} \\
(\epsilon _{2}+3z_{1}-z_{5}-\stackrel{\_}{z}_{5}-\varepsilon z_{4})A(k,k+1)
&=&A(k-1,k+1)+A(k,k+2)  \nonumber \\
&&+\varepsilon \stackrel{\_}{z}_{6}B(k+1)+\varepsilon z_{6}B(k)
\label{eq3.11} \\
(\epsilon _{2}+2z_{1}-z_{7}-\stackrel{\_}{z}_{7})B(k)
&=&z_{3}B(k-1)+z_{3}B(k+1)  \nonumber \\
&&+\stackrel{\_}{z}_{6}A(k-1,k)+z_{6}A(k,k+1)  \label{eq3.12}
\end{eqnarray}
The parametrization (\ref{eq3.8}) solves the equations (\ref{eq3.10})
provided 
\begin{equation}
\epsilon _{2}=-4z_{1}+2z_{5}+2\stackrel{\_}{z}_{5}+\xi _{1}+\xi
_{1}^{-1}+\xi _{2}+\xi _{2}^{-1}  \label{eq3.13}
\end{equation}
It immediately follows that the eigenvalues of $H$ are a sum of single
pseudoparticle energies.

The parametrization of $B(k)$ can now be determined in the following way:
Subtracting Eq.(\ref{eq3.11}) from Eq.(\ref{eq3.10}) for $k_{1}=k,k_{2}=k+1$%
, we get a \ {\em meeting equation} 
\begin{equation}
\varepsilon \stackrel{\_}{z}_{6}B(k+1)+\varepsilon
z_{6}B(k)=A(k,k)+A(k+1,k+1)-(z_{1}+\varepsilon z_{4}-z_{5}-\stackrel{\_}{z}%
_{5})A(k,k+1)  \label{eq3.14}
\end{equation}
Now we extend the parametrization (\ref{eq3.8}) to $k_{1}=k_{2}$ in order to
get a parametrization for the wavefunction $B(k)$ : 
\begin{equation}
B(k)=B(\xi _{1}\xi _{2})^{k}  \label{eq3.15}
\end{equation}
which solves the meeting equation (\ref{eq3.14}) provided 
\begin{eqnarray}
B &=&\varepsilon \frac{1+\xi _{1}\xi _{2}-\Delta _{1}\xi _{2}}{z_{6}+%
\stackrel{\_}{z}_{6}\xi _{1}\xi _{2}}A_{12}+\varepsilon \frac{1+\xi _{1}\xi
_{2}-\Delta _{1}\xi _{1}}{z_{6}+\stackrel{\_}{z}_{6}\xi _{1}\xi _{2}}A_{21} 
\nonumber \\
\Delta _{1} &=&z_{1}+\varepsilon z_{4}-z_{5}-\stackrel{\_}{z}_{5}
\label{eq3.16}
\end{eqnarray}
These relations tell us that the pseudoparticle of the type $\left|
k[-]\right\rangle $ behaves as the two pseudoparticles $\left|
k_{1}[0]\right\rangle $ and $\left| k_{2}[0]\right\rangle $ at the same site 
$k$ and its parametrization follows as the plane waves of particles $\left|
k_{i}[0]\right\rangle $ multiplied by the weight function $B=B(\xi _{1},\xi
_{2})$.

Now substituting (\ref{eq3.8}), (\ref{eq3.13}) and (\ref{eq3.15}) into the
equation (\ref{eq3.12}) we find the phase shift of two pseudoparticles. 
\begin{equation}
\frac{A_{21}}{A_{12}}\equiv \Phi _{12}=-\frac{(1+\xi _{1}\xi
_{2})^{2}-(1+\xi _{1}\xi _{2})(\Delta _{2}\xi _{1}+\Delta _{3}\xi
_{2})+\Delta _{4}\xi _{1}\xi _{2}+\Delta _{5}\xi _{2}^{2}}{(1+\xi _{1}\xi
_{2})^{2}-(1+\xi _{1}\xi _{2})(\Delta _{2}\xi _{2}+\Delta _{3}\xi
_{1})+\Delta _{4}\xi _{1}\xi _{2}+\Delta _{5}\xi _{1}^{2}}  \label{eq3.17}
\end{equation}
where 
\begin{eqnarray}
\Delta _{2} &=&\frac{1}{z_{3}}  \nonumber \\
\Delta _{3} &=&\frac{1}{z_{3}}+\frac{\varepsilon }{z_{3}}(z_{3}z_{4}-z_{6}%
\stackrel{\_}{z}_{6})+(z_{1}-z_{5}-\stackrel{\_}{z}_{5})  \nonumber \\
\Delta _{4} &=&\frac{1}{z_{3}}(\varepsilon z_{4}+z_{7}+\stackrel{\_}{z}_{7})+%
\frac{3}{z_{3}}(z_{1}-\stackrel{\_}{z}_{5}-z_{5})-2  \nonumber \\
\Delta _{5} &=&\frac{1}{z_{3}}(\varepsilon z_{4}+z_{1}-\stackrel{\_}{z}%
_{5}-z_{5})  \label{eq3.18}
\end{eqnarray}
Combining this result with the periodic relations (\ref{eq3.9}) and using (%
\ref{HZF}-\ref{HOSP}) we arrive to the {\small BA} equations in ${\cal H}%
_{2}^{(N)}$ for each model: 
\begin{equation}
\xi _{2}^{N}=-\left( \frac{1+\xi _{1}\xi _{2}+\xi _{1}+\xi _{2}-(\Delta
+2)\xi _{2}}{1+\xi _{1}\xi _{2}+\xi _{1}+\xi _{2}-(\Delta +2)\xi _{1}}\right)
\label{eq3.19}
\end{equation}
for the {\small ZF} model, 
\begin{equation}
\xi _{2}^{N}=-\left( \frac{1+\xi _{1}\xi _{2}-\Delta \xi _{2}}{1+\xi _{1}\xi
_{2}-\Delta \xi _{1}}\right) \left( \frac{1+\xi _{1}\xi _{2}-\xi _{1}-\xi
_{2}+(\Delta -2)\xi _{1}}{1+\xi _{1}\xi _{2}-\xi _{1}-\xi _{2}+(\Delta
-2)\xi _{2}}\right)  \label{eq3.20}
\end{equation}
for the {\small IK} model \ and for the $osp(1|2)$ model we get\ 
\begin{equation}
\xi _{2}^{N}=-\left( \frac{1+\xi _{1}\xi _{2}-\Delta \xi _{2}}{1+\xi _{1}\xi
_{2}-\Delta \xi _{1}}\right) \left( \frac{1+\xi _{1}\xi _{2}+\xi _{1}+\xi
_{2}-(\Delta +2)\xi _{1}}{1+\xi _{1}\xi _{2}+\xi _{1}+\xi _{2}-(\Delta
+2)\xi _{2}}\right)  \label{eq3.21}
\end{equation}
where 
\begin{equation}
\Delta =2\cosh 2\eta ,\qquad {\rm and}\qquad (\xi _{1}\xi _{2})^{N}=1.
\label{eq3.22}
\end{equation}

\subsection{Sector r=3}

Now the Hilbert space is ${\cal H}_{3}^{(N)}$ where there are $N(N-1)(N-2)/6$
states of the type $\left| k_{1}[0],k_{2}[0],k_{3}[0]\right\rangle $, $%
N(N-1)/2$ states of the type $\left| k_{1}[-],k_{2}[0]\right\rangle $ and $%
N(N-1)/2$ states of the type $\left| k_{1}[0],k_{2}[-]\right\rangle $ . We
seek eigenfunctions of the form 
\begin{eqnarray}
\Psi _{3} &=&\sum_{k_{1}<k_{2}<k_{3}}A(k_{1},k_{2},k_{3})\left|
k_{1}[0],k_{2}[0],k_{3}[0]\right\rangle  \nonumber \\
&&+\sum_{k<k}\left\{ B_{1}(k_{1},k_{2})\left| k_{1}[-],k_{2}[0]\right\rangle
+B_{2}(k_{1},k_{2})\left| k_{1}[0],k_{2}[-]\right\rangle \right\}
\label{eq3.23}
\end{eqnarray}
Periodic boundary conditions read now 
\begin{equation}
A(k_{2},k_{3},N+k_{1})=A(k_{1},k_{2},k_{3}),\quad
B_{2}(k_{2},N+k_{1})=B_{1}(k_{1},k_{2}),  \label{eq3.24}
\end{equation}
Again, the wavefunctions $A(k_{1},k_{2},k_{3})$ can be parametrized by the
superposition of plane waves 
\begin{eqnarray}
A(k_{1},k_{2},k_{3}) &=&A_{123}\xi _{1}^{k_{1}}\xi _{2}^{k_{2}}\xi
_{3}^{k_{3}}+A_{132}\xi _{1}^{k_{1}}\xi _{2}^{k_{3}}\xi
_{3}^{k_{2}}+A_{213}\xi _{1}^{k_{2}}\xi _{2}^{k_{1}}\xi _{3}^{k_{3}} 
\nonumber \\
&&+A_{231}\xi _{1}^{k_{2}}\xi _{2}^{k_{3}}\xi _{3}^{k_{1}}+A_{312}\xi
_{1}^{k_{3}}\xi _{2}^{k_{1}}\xi _{3}^{k_{2}}+A_{123}\xi _{1}^{k_{3}}\xi
_{2}^{k_{2}}\xi _{3}^{k_{1}}  \label{eq3.25}
\end{eqnarray}
which satisfy the periodic boundary condition provided 
\begin{equation}
\frac{A_{231}}{A_{123}}=\frac{A_{321}}{A_{213}}=\xi _{3}^{N},\quad \frac{%
A_{213}}{A_{132}}=\frac{A_{312}}{A_{231}}=\xi _{2}^{N},\quad \frac{A_{123}}{%
A_{312}}=\frac{A_{132}}{A_{321}}=\xi _{1}^{N}  \label{eq3.26}
\end{equation}
These relations tell us that the interchange of two pseudoparticles is
independent of the position of the third pseudoparticle. Using $S$-matrix
language, this locality of the interactions is equivalent to the
factorization property of the $S$-matrix, according to which the scattering
amplitude of three particles factorizes into a product of three two-point $S$%
-matrices.

Action of $H$ on these eigenstates gives the following set of coupled
equations for $A(k_{1},k_{2},k_{3})$ and $B_{i}(k_{1},k_{2})$, $i=1,2$: 
\begin{eqnarray}
&&(\epsilon _{3}+6z_{1}-3z_{5}-3\stackrel{\_}{z}_{5})A(k_{1},k_{2},k_{3}) 
\nonumber \\
&=&A(k_{1}-1,k_{2},k_{3})+A(k_{1}+1,k_{2},k_{3})  \nonumber \\
&&+A(k_{1},k_{2}-1,k_{3})+A(k_{1},k_{2}+1,k_{3})  \nonumber \\
&&+A(k_{1},k_{2},k_{3}-1)+A(k_{1},k_{2},k_{3}+1)  \label{eq3.27}
\end{eqnarray}
These equations show us the action of $H$ in configurations of the Hilbert
space ${\cal H}_{3}^{(N)}$ for which the three pseudoparticles ( $\left|
k_{i}[0]\right\rangle $, $\ i=1,2,3$ ) are separated. We already know that
they are satisfied with the plane wave parametrization (\ref{eq3.25}) and
that 
\begin{equation}
\epsilon _{3}=\sum_{j=1}^{3}(-2z_{1}+z_{5}+\stackrel{\_}{z}_{5}+\xi _{j}+\xi
_{j}^{-1}).  \label{eq3.28}
\end{equation}
For configurations where two pseudoparticles are neighbors at $k_{1}$ and
the third pseudoparticles is at $k_{2}>k_{1}+2$, $H$ gives us the following
equations 
\begin{eqnarray}
&&(\epsilon _{3}+5z_{1}-2z_{5}-2\stackrel{\_}{z}_{5}-\varepsilon
z_{4})A(k_{1},k_{1}+1,k_{2})  \nonumber \\
&=&A(k_{1}-1,k_{1}+1,k_{2})+A(k_{1},k_{1}+2,k_{2})  \nonumber \\
&&+A(k_{1},k_{1}+1,k_{2}-1)+A(k_{1},k_{1}+1,k_{2}+1)  \nonumber \\
&&+\varepsilon \stackrel{\_}{z}_{6}B_{1}(k_{1}+1,k_{2})+\varepsilon
z_{6}B_{1}(k_{1},k_{2})  \label{eq3.29}
\end{eqnarray}
and a similar set of equations coupling $B_{2}(k_{1},k_{2})$ and $%
A(k_{1},k_{2},k_{3})$, which correspond to the meet of two pseudoparticle on
the right hand side of the third pseudparticle.

Comparing the Eq.(\ref{eq3.29}) with the Eq.(\ref{eq3.27}) we get a
consistency equation 
\begin{eqnarray}
\varepsilon \stackrel{\_}{z}_{6}B_{1}(k_{1}+1,k_{2})+\varepsilon
z_{6}B_{1}(k_{1},k_{2}) &=&A(k_{1},k_{1},k_{2})+A(k_{1}+1,k_{1}+1,k_{2}) 
\nonumber \\
&&-\Delta _{1}\ A(k_{1},k_{1}+1,k_{2})  \label{eq3.40}
\end{eqnarray}
Similarly, for the right hand side meeting we get 
\begin{eqnarray}
\varepsilon \stackrel{\_}{z}_{6}B_{2}(k_{1},k_{2}+1)+\varepsilon
z_{6}B_{2}(k_{1},k_{2}) &=&A(k_{1},k_{2},k_{2})+A(k_{1},k_{2}+1,k_{2}+1) 
\nonumber \\
&&-\Delta _{1}A(k_{1},k_{1}+1,k_{2})  \label{eq3.41}
\end{eqnarray}

These consistency equations are solved by the following parametrization of
the wavefunctions $B_{i}(k_{1},k_{2})$, $i=1,2$. 
\begin{eqnarray}
B_{1}(k_{1},k_{2}) &=&B_{11}(\xi _{1}\xi _{2})^{k_{1}}\xi
_{3}^{k_{2}}+B_{12}(\xi _{1}\xi _{3})^{k_{1}}\xi _{2}^{k_{2}}+B_{13}(\xi
_{2}\xi _{3})^{k_{1}}\xi _{1}^{k_{2}}  \nonumber \\
B_{2}(k_{1},k_{2}) &=&B_{21}\xi _{1}^{k_{1}}(\xi _{2}\xi
_{3})^{k_{2}}+B_{22}\xi _{2}^{k_{1}}(\xi _{1}\xi _{3})^{k_{2}}+B_{23}\xi
_{3}^{k_{1}}(\xi _{1}\xi _{2})^{k_{2}}  \label{eq3.42}
\end{eqnarray}
which satisfy the periodic boundary condition provide 
\begin{equation}
B_{21}=\xi _{1}^{N}B_{13},\quad B_{22}=\xi _{2}^{N}B_{12},\quad B_{23}=\xi
_{3}^{N}B_{11}  \label{eq3.43}
\end{equation}
Moreover, the weight functions $B_{1i}$ and $B_{2i},$ $i=1,2,3$ are
determined 
\begin{equation}
\begin{array}{lll}
B_{11}=F_{12}A_{123}+F_{21}A_{213} & , & B_{21}=F_{23}A_{123}+F_{32}A_{132}
\\ 
B_{12}=F_{13}A_{132}+F_{31}A_{231} & , & B_{22}=F_{13}A_{213}+F_{31}A_{312}
\\ 
B_{13}=F_{23}A_{312}+F_{23}A_{321} & , & B_{23}=F_{12}A_{231}+F_{21}A_{321}
\end{array}
\label{eq3.44}
\end{equation}
where 
\begin{equation}
F_{ab}=\varepsilon \frac{1+\xi _{a}\xi _{b}-\Delta _{1}\xi _{b}}{z_{6}+%
\stackrel{\_}{z}_{6}\xi _{a}\xi _{b}},\quad a\neq b=1,2,3  \label{eq3.45}
\end{equation}
Substituting these relations into the eigenvalue equations (\ref{eq3.29}) we
obtain the phase shift of two pseudoparticles 
\begin{equation}
\frac{A_{123}}{A_{213}}=\frac{A_{231}}{A_{321}}=\Phi _{12},\quad \frac{%
A_{132}}{A_{231}}=\frac{A_{213}}{A_{312}}=\Phi _{13},\quad \frac{A_{312}}{%
A_{321}}=\frac{A_{123}}{A_{132}}=\Phi _{23}  \label{eq3.46}
\end{equation}
where 
\begin{equation}
\Phi _{ab}=-\frac{(1+\xi _{a}\xi _{b})^{2}-(1+\xi _{a}\xi _{b})(\Delta
_{2}\xi _{a}+\Delta _{3}\xi _{b})+\Delta _{4}\xi _{a}\xi _{b}+\Delta _{5}\xi
_{b}^{2}}{(1+\xi _{a}\xi _{b})^{2}-(1+\xi _{a}\xi _{b})(\Delta _{2}\xi
_{b}+\Delta _{3}\xi _{a})+\Delta _{4}\xi _{a}\xi _{b}+\Delta _{5}\xi _{a}^{2}%
},\quad a\neq b=1,2,3  \label{eq3.47}
\end{equation}
and the $\Delta _{i},\ i=1,2,3,4$ are given by (\ref{eq3.16}) and (\ref
{eq3.18})

Next, when the three pseudoparticles are neighbors we have the following
eigenvalue equations 
\begin{eqnarray}
&&(\epsilon _{3}+4z_{1}-z_{5}-\stackrel{\_}{z}_{5}-2\varepsilon
z_{4})A(k,k+1,k+2)  \nonumber \\
&=&A(k-1,k+1,k+2)+A(k,k+1,k+3)  \nonumber \\
&&+\varepsilon \stackrel{\_}{z}_{6}B_{1}(k+1,k+2)+\varepsilon
z_{6}B_{1}(k,k+2)  \nonumber \\
&&+\varepsilon \stackrel{\_}{z}_{6}B_{2}(k,k+2)+\varepsilon z_{6}B_{2}(k,k+1)
\label{eq3.48}
\end{eqnarray}
which are automatically satisfied by the above parametrizations.

In addition to this equations we also have to consider the equations for
configurations where the pseudoparticle of the type $\left|
k[-]\right\rangle $ and the pseudoparticle $\left| k[0]\right\rangle $ are
separated: 
\begin{eqnarray}
&&(\epsilon _{3}+4z_{1}-z_{5}-\stackrel{\_}{z}_{5}-\stackrel{\_}{z}%
_{7}-z_{7})B_{1}(k_{1},k_{2})  \nonumber \\
&=&B_{1}(k_{1},k_{2}-1)+B_{1}(k_{1},k_{2}+1)  \nonumber \\
&&+z_{3}B_{1}(k_{1}-1,k_{2})+z_{3}B_{1}(k_{1}+1,k_{2})  \nonumber \\
&&+\stackrel{\_}{z}_{6}A(k_{1}-1,k_{1},k_{2})+z_{6}A(k_{1}+1,k_{1},k_{2})
\label{eq3.49}
\end{eqnarray}
and similar set of eigenvalue equation involving $B_{2}(k_{1},k_{2})$, which
corresponds to configurations with the pseudoparticle $\left|
k_{2}[-]\right\rangle $ on the right side hand of the pseudoparticle $\left|
k_{1}[0]\right\rangle $.

These equations are also satisfied by the above parametrizations. This
statement was already expected since at this point we always have a {\em far}
particle as a {\em viewer}. Therefore, no appeared configurations different
from those presented in the sector $r=2$.

Finally, the action of $H$ on configurations where the two different
pseudoparticles are neighbors results in two more eigenvalue equations: 
\begin{eqnarray}
(\epsilon _{3}+3z_{1}-\stackrel{\_}{z}_{7}-2\stackrel{\_}{z}%
_{5})B_{1}(k,k+1) &=&B_{1}(k,k+2)+B_{2}(k,k+1)  \nonumber \\
&&+z_{3}B_{1}(k-1,k+1)+\stackrel{\_}{z}_{6}A(k-1,k,k+1)  \nonumber \\
&&  \label{eq3.50}
\end{eqnarray}
and 
\begin{eqnarray}
(\epsilon _{3}+3z_{1}-z_{7}-2z_{5})B_{2}(k,k+1)
&=&B_{2}(k-1,k+1)+B_{1}(k,k+1)  \nonumber \\
&&+z_{3}B_{2}(k,k+2)+z_{6}A(k,k+1,k+2)  \label{eq3.51}
\end{eqnarray}

Substituting the wavefunctions parametrizations for $A(k_{1},k_{2},k_{3})$
and $B_{i}(k_{1},k_{2})$ into the equations (\ref{eq3.50}) and (\ref{eq3.51}%
) and using the relations (\ref{eq3.26}) and (\ref{eq3.46}) one can verify
that they are indeed satisfied. These results tell us that the meeting of
the pseudoparticle $\left| k[-]\right\rangle $ with the pseudoparticle $%
\left| k[0]\right\rangle $ can be versed as a meeting of three
pseudoparticle $\left| k[0]\right\rangle $.

Compounding (\ref{eq3.46}) with the periodic boundary conditions (\ref
{eq3.26}) we arrive to the {\small BA} equations for the sector $r=3$%
\begin{equation}
\xi _{a}^{N}=\prod_{b\neq a}^{3}\Phi _{ab},\quad a=1,2,3  \label{eq3.52}
\end{equation}
which expresses the factorization of the three pseudoparticle phase shift
into the product of two pseudoparticle phase shifts.

\subsection{General sector}

The above results can now be generalized. First we observe that in the
sector $r>3$ there is no additional meeting conditions. For example in the
sector $r=4$ there is a meeting of two pseudoparticles of the type $\left|
k[-]\right\rangle $. Nevertheless, we know that the state $\left|
k[-]\right\rangle $ is parametrized as two states $\left| k[0]\right\rangle $
at the same site and we have verified that the meeting of two
pseudoparticles $\left| k[-]\right\rangle $ behaves as the meeting of four
pseudoparticles $\left| k[0]\right\rangle $ whose phase shift factorizes in
a product of two pseudoparticle phase shifts.

In a generic sector $r$ we build eigenstates of $H$ out of translational
invariant products of $N_{0}$ one-pseudoparticle eigenstates $\left|
k[0]\right\rangle $ and $N_{-}$ two-pseudoparticle eigenstates $\left|
k[-]\right\rangle $, such that $r=N_{0}+2N_{-}$ . These eigenstates are
obtained by superposition of terms of the form 
\begin{equation}
\left| \phi _{r}\right\rangle =\left| 0\right\rangle \times \left| \phi
_{r-1}\right\rangle +\left| -\right\rangle \times \left| \phi
_{r-2}\right\rangle  \label{eq3.53}
\end{equation}
with $\left| \phi _{0}\right\rangle =1$, $\left| \phi _{1}\right\rangle
=\left| 0\right\rangle $. The corresponding eigenvalue is a sum of single
one-particle energies 
\begin{equation}
E_{r}=Nz_{1}+\sum_{a=1}^{r}(-2z_{1}+z_{5}+\stackrel{\_}{z}_{5}+\xi _{a}+\xi
_{a}^{-1})  \label{eq3.54}
\end{equation}
being $\xi _{a}$ solutions of the {\small BA} equations 
\begin{equation}
\xi _{a}^{N}=\prod_{b\neq a=1}^{r}\Phi _{ab},\quad a=1,2,...,r
\label{eq3.55}
\end{equation}
where 
\begin{eqnarray}
\Phi _{ab} &=&-\left( \frac{1+\xi _{a}\xi _{b}+\xi _{a}+\xi _{b}-(\Delta
+2)\xi _{b}}{1+\xi _{a}\xi _{b}+\xi _{a}+\xi _{b}-(\Delta +2)\xi _{a}}%
\right) ,  \nonumber \\
\ a,b &=&1,2,...,r.\quad \Delta =2\cosh 2\eta  \label{eq3.56}
\end{eqnarray}
for the {\small ZF} model, 
\begin{eqnarray}
\Phi _{ab} &=&-\left( \frac{1+\xi _{a}\xi _{b}-\Delta \xi _{b}}{1+\xi
_{a}\xi _{b}-\Delta \xi _{a}}\right) \left( \frac{1+\xi _{a}\xi _{b}-\xi
_{a}-\xi _{b}+(\Delta -2)\xi _{a}}{1+\xi _{a}\xi _{b}-\xi _{a}-\xi
_{b}+(\Delta -2)\xi _{b}}\right)  \nonumber \\
a,b &=&1,2,...,r.\quad \Delta =2\cosh 2\eta  \label{eq3.57}
\end{eqnarray}
for the {\small IK} model and 
\begin{eqnarray}
\Phi _{ab} &=&-\left( \frac{1+\xi _{a}\xi _{b}-\Delta \xi _{b}}{1+\xi
_{a}\xi _{b}-\Delta \xi _{a}}\right) \left( \frac{1+\xi _{a}\xi _{b}+\xi
_{a}+\xi _{b}-(\Delta +2)\xi _{a}}{1+\xi _{a}\xi _{b}+\xi _{a}+\xi
_{b}-(\Delta +2)\xi _{b}}\right)  \nonumber \\
a,b &=&1,2,...,r.\quad \Delta =2\cosh 2\eta  \label{eq3.58}
\end{eqnarray}
for the $osp(1|2)$ model.

\section{The Algebraic Bethe Ansatz}

In the previous section we have considered the problem of diagonalization of
a one-dimensional spin chain Hamiltonian using the Coordinate {\small BA}.
Let us now turn to two dimensional classical statistical systems on a
lattice.

Let us consider a regular lattice with $N$ columns and $N^{\prime }$ rows. A
physical state on this lattice is defined by the assignment of a {\em state
variable }to each lattice edge. If one takes the horizontal direction as
space and the vertical one as time, the transfer matrix plays the role of a
discrete evolution operator acting on the Hilbert space ${\cal H}^{(N)}$
spanned by the {\em row states} which are defined by the set of vertical
link variables on the same row. Thus, the transfer matrix elements can be
understood as the transition probability of the one row state to project on
\ the consecutive one after a unit of time.

The main problem now is the diagonalization of the transfer matrix of the
lattice system. To do this we request the Algebraic {\small BA}.

Again, we start with the graded formulation such that the additional signs
are represented by $\varepsilon =-1$. Taking $\varepsilon =1$ we recover the
non graded cases.

We recall some basic relations of the graded quantum inverse scattering
method. For us the basic object will be the $R$ matrix (\ref{eq2.6}), which
satisfies $R(0,\eta )=\rho (\eta )\ {\cal I}$, where $\rho _{{\rm ZF}}(\eta
)=\sinh \eta \sinh 2\eta ,\quad \rho _{{\rm IK}}(\eta )=-\sinh 5\eta +\sinh
\eta \ $ and\ $\rho _{{\rm osp}}(\eta )=\sinh 2\eta \sinh 3\eta $.

A quantum integrable system is characterized by monodromy matrix $T(\lambda
) $ satisfying the equation 
\begin{equation}
R(\lambda -\mu )\left[ T(\lambda )\stackrel{s}{\otimes }T(\mu )\right] =%
\left[ T(\mu )\stackrel{s}{\otimes }T(\lambda )\right] R(\lambda -\mu )
\label{eq4.1}
\end{equation}
whose consistency is guaranteed by the {\small YB} equation (\ref{eq2.5}). $%
T(\lambda )$ is a matrix in the space $V$ with matrix elements that are
operators on the states of the quantum system (the quantum space, which will
also be the space $V$). The space $V$ is called auxiliary space of $%
T(\lambda )$. An example of a monodromy matrix is the matrix ${\cal P}R$,
this follows directly from (\ref{eq4.1}).

The simplest monodromies have become known as ${\cal L}$ operators, the Lax
operator, and the monodromy operator $T(\lambda )$ is defined as an ordered
product of Lax operators on all sites of the lattice: 
\begin{equation}
T(\lambda )={\cal L}_{N}(\lambda ){\cal L}_{N-1}(\lambda )\cdots {\cal L}%
_{1}(\lambda ).  \label{eq4.2}
\end{equation}
The Lax operator on the $n^{th}$ quantum space is given the graded
permutation of (\ref{eq2.6}): 
\begin{eqnarray}
{\cal L}_{n}(\lambda ) &=&\left( 
\begin{array}{lllllllll}
x_{1} & 0 & 0 & 0 & \ \ 0 & 0 & \ \ 0 & 0 & 0 \\ 
0 & x_{2} & 0 & x_{5} & \ \ 0 & 0 & \ \ 0 & 0 & 0 \\ 
0 & 0 & x_{3} & 0 & x_{6} & 0 & \ x_{7} & 0 & 0 \\ 
0 & y_{5} & 0 & x_{2} & \ \ 0 & 0 & \ \ 0 & 0 & 0 \\ 
0 & 0 & y_{6} & 0 & x_{4} & 0 & x_{6} & 0 & 0 \\ 
0 & 0 & 0 & 0 & \ \ 0 & x_{2} & \ \ 0 & x_{5} & 0 \\ 
0 & 0 & y_{7} & 0 & \ y_{6} & 0 & \ x_{3} & 0 & 0 \\ 
0 & 0 & 0 & 0 & \ \ 0 & y_{5} & \ \ 0 & x_{2} & 0 \\ 
0 & 0 & 0 & 0 & \ \ 0 & 0 & \ \ 0 & 0 & x_{1}
\end{array}
\right)  \nonumber \\
&=&\left( 
\begin{array}{lll}
L_{11}^{(n)}(\lambda ) & L_{12}^{(n)}(\lambda ) & L_{13}^{(n)}(\lambda ) \\ 
L_{21}^{(n)}(\lambda ) & L_{22}^{(n)}(\lambda ) & L_{23}^{(n)}(\lambda ) \\ 
L_{31}^{(n)}(\lambda ) & L_{32}^{(n)}(\lambda ) & L_{33}^{(n)}(\lambda )
\end{array}
\right)  \label{eq4.3}
\end{eqnarray}
Note that $L_{ij}^{(n)}(\lambda ),\ i,j=1,2,3$ are $3$ by $3$ matrices
acting on the $n^{th}$ site of the lattice. It means that the monodromy
matrix has the form 
\begin{equation}
T(\lambda )=\left( 
\begin{array}{lll}
T_{11}(\lambda ) & T_{12}(\lambda ) & T_{13}(\lambda ) \\ 
T_{21}(\lambda ) & T_{22}(\lambda ) & T_{23}(\lambda ) \\ 
T_{31}(\lambda ) & T_{32}(\lambda ) & T_{33}(\lambda )
\end{array}
\right) =\left( 
\begin{array}{lll}
A_{1}(\lambda ) & B_{1}(\lambda ) & B_{2}(\lambda ) \\ 
C_{1}(\lambda ) & A_{2}(\lambda ) & B_{3}(\lambda ) \\ 
C_{2}(\lambda ) & C_{3}(\lambda ) & A_{3}(\lambda )
\end{array}
\right)  \label{eq4.4}
\end{equation}
where 
\begin{eqnarray}
T_{ij}(\lambda ) &=&\sum_{k_{1},...,k_{N-1}=1}^{3}L_{ik_{1}}^{(N)}(\lambda )%
\stackrel{s}{\otimes }L_{k_{1}k_{2}}^{(N-1)}(\lambda )\stackrel{s}{\otimes }%
\cdots \stackrel{s}{\otimes }L_{k_{N-1}j}^{(1)}(\lambda )  \nonumber \\
i,j &=&1,2,3.  \label{eq4.5}
\end{eqnarray}

The vector $\left| 0\right\rangle $ in the quantum space of the monodromy
matrix $T(\lambda )$ that is annihilated by the operators $T_{ij}(\lambda )$%
, $i>j$ ($C_{k}(\lambda )$ operators, $k=1,2,3$) and is an eigenvector for
the operators $T_{ii}(\lambda )$ ( $A_{k}(\lambda )$ operators, $k=1,2,3$)
is called a highest vector of the monodromy matrix $T(\lambda )$.

The transfer matrix $\tau (\lambda )$ of the corresponding integrable spin
model is given by the supertrace of the monodromy matrix in the space $V$, 
{\rm Str}$T(\lambda )$. It is the generating function of the family of
commuting operators in terms of which the Hamiltonian of the quantum system
is expressed. 
\begin{equation}
\tau (\lambda )={\rm Str}T(\lambda )=\sum_{i=1}^{3}(-)^{p(i)}\
T_{ii}(\lambda )=A_{1}(\lambda )+\varepsilon A_{2}(\lambda )+A_{3}(\lambda )
\label{eq4.6}
\end{equation}
In particular, the Hamiltonians (\ref{eq2.10}) can also be derived by the
well-known relation 
\begin{equation}
H=\alpha \ \frac{\partial }{\partial \lambda }\left( \ln \tau (\lambda
)\right) _{\lambda =0}  \label{eq4.7}
\end{equation}

A detailed exposition of the graded quantum inverse scattering method can be
found in reference \cite{EK}.

In this section we will derive the {\small BA} equations of $19$-vertex
models presented in Section $2$ using the Algebraic {\small BA} developed by
Tarasov \cite{TA} and recently generalized by Martins and Ramos \cite{MR}.
To do this we need of the commutation relations for entries of the monodromy
matrix which are derived from the fundamental relation (\ref{eq4.1}). Here
these commutation relations do not share a common structure. Therefore, we
only write some of them in the text and recall (\ref{eq4.1}) to get the
remaining ones.

First of all , let us observe that for each row state one can define the
magnon number operator which commutes with the transfer matrix of the models 
\begin{equation}
\lbrack \tau (\lambda ),M]=0,\quad M=\sum_{k=1}^{N}M_{k},\qquad M_{k}=\left( 
\begin{array}{lll}
0 & 0 & 0 \\ 
0 & 1 & 0 \\ 
0 & 0 & 2
\end{array}
\right) ,  \label{eq4.8}
\end{equation}
This is the analog of the operator $S_{T}^{z}$ used in the previous section
and the relation between $M$ and the spin total $S_{T}^{z}$ is simply $%
M=N-S_{T}^{z}$. Once again, the Hilbert space can be broken down into
sectors ${\cal H}_{M}^{(N)}$. In each of these sectors, the transfer matrix
can be diagonalized independently, $\tau (\lambda )\Psi _{M}=\Lambda
_{M}\Psi _{M}$ . We will now start to diagonalize $\tau (\lambda )$ in every
sector:

\subsection{Sector $M=0$}

Let us consider the highest vector of the monodromy matrix $T(\lambda )$ in
a lattice of $N$ sites as the even (bosonic) completely unoccupied state 
\begin{equation}
\Psi _{0}\equiv \left| 0\right\rangle =\otimes _{k=1}^{N}\left( 
\begin{array}{l}
1 \\ 
0 \\ 
0
\end{array}
\right) _{k}  \label{eq4.9}
\end{equation}
It is the only state in the sector with $M=0$. Using (\ref{eq4.5}) we can
compute the action of the matrix elements of $T(\lambda )$ on this reference
state: 
\begin{eqnarray}
\quad A_{1}(\lambda )\left| 0\right\rangle &=&x_{1}^{N}(\lambda )\left|
0\right\rangle ,\quad A_{2}(\lambda )\left| 0\right\rangle
=x_{2}^{N}(\lambda )\left| 0\right\rangle ,\quad A_{3}(\lambda )\left|
0\right\rangle =x_{3}^{N}(\lambda )\left| 0\right\rangle  \nonumber \\
\quad C_{k}(\lambda )\left| 0\right\rangle &=&0,\quad B_{k}(\lambda )\left|
0\right\rangle \neq \{0,\left| 0\right\rangle \},\quad k=1,2,3
\label{eq4.10}
\end{eqnarray}
Therefore in the sector $M=0$ , $\Psi _{0}$ is the eigenstate of $\tau
(\lambda )=A_{1}(\lambda )+\varepsilon A_{2}(\lambda )+A_{3}(\lambda )$ with
eigenvalue 
\begin{equation}
\Lambda _{0}(\lambda )=x_{1}^{N}(\lambda )+\varepsilon x_{2}^{N}(\lambda
)+x_{3}^{N}(\lambda )  \label{eq4.11}
\end{equation}
Here we observe that the action of the operators $B_{1}(\lambda )$, $%
B_{2}(\lambda )$ and $B_{3}(\lambda )$ on the reference state will give us
new states which lie in sectors with $M\neq 0$.

\subsection{Sector $M=1$}

In this sector we have the states $B_{1}(\lambda )\left| 0\right\rangle $
and $B_{3}(\lambda )\left| 0\right\rangle $. Since $B_{3}(\lambda )\left|
0\right\rangle \propto B_{1}(\lambda )\left| 0\right\rangle $, we seek
eigenstate of the form 
\begin{equation}
\Psi _{1}(\lambda _{1})=B_{1}(\lambda _{1})\left| 0\right\rangle .
\label{eq4.12}
\end{equation}

The action of the operator $\tau (\lambda )$ on this state can be computed
with aid of the following commutation relations 
\begin{eqnarray}
A_{1}(\lambda )B_{1}(\mu ) &=&z(\mu -\lambda )B_{1}(\mu )A_{1}(\lambda )-%
\frac{x_{5}(\mu -\lambda )}{x_{2}(\mu -\lambda )}B_{1}(\lambda )A_{1}(\mu )
\label{eq4.13a} \\
A_{2}(\lambda )B_{1}(\mu ) &=&\varepsilon \frac{z(\lambda -\mu )}{\omega
(\lambda -\mu )}B_{1}(\mu )A_{2}(\lambda )-\frac{z(\lambda -\mu )}{\omega
(\lambda -\mu )}\frac{1}{y(\mu -\lambda )}B_{2}(\mu )C_{1}(\lambda ) 
\nonumber \\
&&-\varepsilon \frac{y_{5}(\lambda -\mu )}{x_{2}(\lambda -\mu )}%
B_{1}(\lambda )A_{2}(\mu )+\frac{y_{5}(\lambda -\mu )}{x_{2}(\lambda -\mu )}%
\frac{1}{y(\lambda -\mu )}B_{2}(\lambda )C_{1}(\mu )  \nonumber \\
&&+\frac{1}{y(\lambda -\mu )}B_{3}(\lambda )A_{1}(\mu ) \\
A_{3}(\lambda )B_{1}(\mu ) &=&\frac{x_{2}(\lambda -\mu )}{x_{3}(\lambda -\mu
)}B_{1}(\mu )A_{3}(\lambda )-\frac{\varepsilon }{y(\lambda -\mu )}%
B_{3}(\lambda )A_{2}(\mu )  \nonumber \\
&&+\frac{x_{5}(\lambda -\mu )}{x_{3}(\lambda -\mu )}B_{2}(\mu )C_{3}(\lambda
)-\frac{y_{7}(\lambda -\mu )}{x_{3}(\lambda -\mu )}B_{2}(\lambda )C_{3}(\mu )
\label{eq4.13}
\end{eqnarray}
The ratio functions which appear in the commutation relations are defined by 
\begin{eqnarray}
z(\lambda ) &=&\frac{x_{1}(\lambda )}{x_{2}(\lambda )},\quad \omega (\lambda
)=\varepsilon \frac{x_{1}(\lambda )x_{3}(\lambda )}{x_{3}(\lambda
)x_{4}(\lambda )-x_{6}(\lambda )y_{6}(\lambda )},  \nonumber \\
\quad y(\lambda ) &=&\frac{x_{3}(\lambda )}{y_{6}(\lambda )},\quad
y(-\lambda )=\varepsilon \frac{x_{3}(\lambda )x_{4}(\lambda )-x_{6}(\lambda
)y_{6}(\lambda )}{x_{7}(\lambda )y_{6}(\lambda )-x_{3}(\lambda
)x_{6}(\lambda )},  \label{eq4.14}
\end{eqnarray}

When $\tau (\lambda )$ act on $\Psi _{1}(\lambda _{1})$ , the corresponding
eigenvalue equation has two unwanted terms: 
\begin{eqnarray}
\tau (\lambda )\Psi _{1}(\lambda _{1}) &=&\left( A_{1}(\lambda )+\varepsilon
A_{2}(\lambda )+A_{3}(\lambda )\right) \Psi _{1}(\lambda _{1})  \nonumber \\
&=&[z(\lambda _{1}-\lambda )x_{1}^{N}(\lambda )+\varepsilon ^{2}\frac{%
z(\lambda -\lambda _{1})}{\omega (\lambda -\lambda _{1})}x_{2}^{N}(\lambda )+%
\frac{x_{2}(\lambda -\lambda _{1})}{x_{3}(\lambda -\lambda _{1})}%
x_{3}^{N}(\lambda )]\Psi _{1}(\lambda _{1})  \nonumber \\
&&-[\frac{x_{5}(\lambda _{1}-\lambda )}{x_{2}(\lambda _{1}-\lambda )}%
x_{1}^{N}(\lambda _{1})+\varepsilon ^{2}\frac{y_{5}(\lambda -\lambda _{1})}{%
x_{2}(\lambda -\lambda _{1})}x_{2}^{N}(\lambda _{1})]B_{1}(\lambda )\left|
0\right\rangle  \nonumber \\
&&+\varepsilon \lbrack \frac{1}{y(\lambda -\lambda _{1})}x_{1}^{N}(\lambda
_{1})-\frac{1}{y(\lambda -\lambda _{1})}x_{2}^{N}(\lambda
_{1})]B_{3}(\lambda )\left| 0\right\rangle  \label{eq4.15}
\end{eqnarray}
From the matrix elements of $19$-vertex models (\ref{RZF}-\ref{ROSP}) we can
see that $x_{5}(\lambda )/x_{2}(\lambda )=-y_{5}(-\lambda )/x_{2}(-\lambda )$%
. Therefore the unwanted terms vanish and $\Psi _{1}(\lambda _{1})$ is
eigenstate of $\tau (\lambda )$ with eigenvalue 
\begin{equation}
\Lambda _{1}(\lambda ,\lambda _{1})=z(\lambda _{1}-\lambda
)x_{1}^{N}(\lambda )+\varepsilon ^{2}\frac{z(\lambda -\lambda _{1})}{\omega
(\lambda -\lambda _{1})}x_{2}^{N}(\lambda )+\frac{x_{2}(\lambda -\lambda
_{1})}{x_{3}(\lambda -\lambda _{1})}x_{3}^{N}(\lambda )  \label{eq4.16}
\end{equation}
provided 
\begin{equation}
\left( \frac{x_{1}(\lambda _{1})}{x_{2}(\lambda _{1})}\right)
^{N}=\varepsilon ^{2}=1  \label{eq4.17}
\end{equation}

\subsection{Sector $M=2$}

In the sector $M=2$, we encounter two linearly independent states $%
B_{1}(\lambda )B_{1}(\mu )\left| 0\right\rangle $ and $B_{2}(\lambda )\left|
0\right\rangle $. (The states $B_{3}B_{3}\left| 0\right\rangle
,B_{1}B_{3}\left| 0\right\rangle $ and $B_{3}B_{1}\left| 0\right\rangle $
also lie in the sector $M=2$ but they are proportional to the state $%
B_{1}B_{1}\left| 0\right\rangle $). We seek eigenstates in the form 
\begin{equation}
\Psi _{2}(\lambda _{1},\lambda _{2})=B_{1}(\lambda _{1})B_{1}(\lambda
_{2})\left| 0\right\rangle +B_{2}(\lambda _{1})\Gamma (\lambda _{1},\lambda
_{2})\left| 0\right\rangle  \label{eq4.18}
\end{equation}
where $\Gamma (\lambda _{1},\lambda _{2})$ is an operator-valued function
which has to be fixed such that $\Psi _{2}(\lambda _{1},\lambda _{2})$ is
unique state in the sector $M=2$.

Here we observe that the operator-valued function $\Gamma (\lambda
_{1},\lambda _{2})$ is the analog of the weight function $B(\xi _{1},\xi
_{2})$ of the Eq.(\ref{eq3.16}).

It was demonstrated in \cite{TA} that $\Psi _{2}(\lambda _{1},\lambda _{2})$
is unique provided it is ordered in a normal way: In general, the
operator-valued function $\Psi _{n}(\lambda _{1},...,\lambda _{n})$ is
composite of normal ordered monomials. A monomial is normally ordered if in
it all elements of the type $B_{i}(\lambda )$ are on the left, and all
elements of the type $C_{j}(\lambda )$ on the right of all elements of the
type $A_{k}(\lambda ).$ Moreover, the elements of one given type having
standard ordering: $T_{i_{1}j_{1}}(\lambda _{1})T_{i_{2}j_{2}}(\lambda
_{2})...T_{i_{n}jn}(\lambda _{n})$. For a given sector $M=$ $n$, $\Psi
_{n}(\lambda _{1},...,\lambda _{n})$ is unique.

From the commutation relation 
\begin{eqnarray}
B_{1}(\lambda )B_{1}(\mu ) &=&\omega (\mu -\lambda )[B_{1}(\mu
)B_{1}(\lambda )-\frac{1}{y(\mu -\lambda )}B_{2}(\mu )A_{1}(\lambda )] 
\nonumber \\
&&+\frac{1}{y(\lambda -\mu )}B_{2}(\lambda )A_{1}(\mu )  \label{eq4.19}
\end{eqnarray}
we can see that (\ref{eq4.19}) will be normally ordered if it satisfies the
following swap condition 
\begin{equation}
\Psi _{2}(\lambda _{2},\lambda _{1})=\omega (\lambda _{1}-\lambda _{2})\Psi
_{2}(\lambda _{1},\lambda _{2})  \label{eq4.20}
\end{equation}
This condition fixes $\Gamma (\lambda _{1},\lambda _{2})$ in Eq.(\ref{eq4.18}%
) and the eigenstate of $\tau (\lambda )$ in the sector $M=2$ has the form 
\begin{equation}
\Psi _{2}(\lambda _{1},\lambda _{2})=B_{1}(\lambda _{1})B_{1}(\lambda
_{2})\left| 0\right\rangle -\frac{1}{y(\lambda _{1}-\lambda _{2})}%
B_{2}(\lambda _{1})A_{1}(\lambda _{2})\left| 0\right\rangle .  \label{eq4.21}
\end{equation}
The action of transfer matrix on the states of the form (\ref{eq4.21}) is
more laborious. In addition to (\ref{eq4.13a}-\ref{eq4.13}) and (\ref{eq4.19}%
) we need appeal to (\ref{eq4.1}) to derive more eight commutation relations 
\begin{eqnarray}
A_{1}(\lambda )B_{2}(\mu ) &=&\frac{x_{1}(\mu -\lambda )}{x_{3}(\mu -\lambda
)}B_{2}(\mu )A_{1}(\lambda )-\frac{x_{7}(\mu -\lambda )}{x_{3}(\mu -\lambda )%
}B_{2}(\lambda )A_{1}(\mu )  \nonumber \\
&&-\varepsilon \frac{x_{6}(\mu -\lambda )}{x_{3}(\mu -\lambda )}%
B_{1}(\lambda )B_{1}(\mu )  \label{eq4.22a} \\
A_{2}(\lambda )B_{2}(\mu ) &=&z(\lambda -\mu )z(\mu -\lambda )B_{2}(\mu
)A_{2}(\lambda )  \nonumber \\
&&+\frac{y_{5}(\lambda -\mu )}{x_{2}(\lambda -\mu )}[B_{1}(\lambda
)B_{3}(\mu )-\varepsilon B_{3}(\lambda )B_{1}(\mu )+\frac{y_{5}(\lambda -\mu
)}{x_{2}(\lambda -\mu )}B_{2}(\lambda )A_{2}(\mu )]  \nonumber \\
&& \\
A_{3}(\lambda )B_{2}(\mu ) &=&\frac{x_{1}(\lambda -\mu )}{x_{3}(\lambda -\mu
)}B_{2}(\mu )A_{3}(\lambda )-\frac{y_{7}(\lambda -\mu )}{x_{3}(\lambda -\mu )%
}B_{2}(\lambda )A_{3}(\mu )  \nonumber \\
&&-\frac{\varepsilon }{y(\lambda -\mu )}B_{3}(\lambda )B_{3}(\mu ) \\
C_{1}(\lambda )B_{1}(\mu ) &=&\varepsilon B_{1}(\mu )C_{1}(\lambda )+\frac{%
y_{5}(\lambda -\mu )}{x_{2}(\lambda -\mu )}[A_{1}(\mu )A_{2}(\lambda
)-A_{1}(\lambda )A_{2}(\mu )] \\
C_{3}(\lambda )B_{1}(\mu ) &=&\varepsilon \frac{x_{4}(\lambda -\mu )}{%
x_{3}(\lambda -\mu )}B_{1}(\mu )C_{3}(\lambda )-\frac{x_{7}(\lambda -\mu )}{%
x_{3}(\lambda -\mu )}B_{1}(\lambda )C_{3}(\mu )  \nonumber \\
&&+\frac{1}{y(\lambda -\mu )}[A_{1}(\mu )A_{3}(\lambda )-A_{2}(\lambda
)A_{2}(\mu )]+\frac{x_{6}(\lambda -\mu )}{x_{3}(\lambda -\mu )}B_{2}(\mu
)C_{2}(\lambda )  \nonumber \\
&& \\
B_{1}(\lambda )B_{2}(\mu ) &=&\frac{1}{z(\lambda -\mu )}B_{2}(\mu
)B_{1}(\lambda )+\frac{y_{5}(\lambda -\mu )}{x_{1}(\lambda -\mu )}B_{1}(\mu
)B_{2}(\lambda ) \\
B_{1}(\lambda )B_{3}(\mu ) &=&\varepsilon B_{3}(\mu )B_{1}(\lambda )-\frac{%
y_{5}(\lambda -\mu )}{x_{2}(\lambda -\mu )}B_{2}(\mu )A_{2}(\lambda )+\frac{%
x_{5}(\lambda -\mu )}{x_{2}(\lambda -\mu )}B_{2}(\lambda )A_{2}(\mu ) 
\nonumber \\
&& \\
B_{2}(\lambda )B_{1}(\mu ) &=&\frac{1}{z(\lambda -\mu )}B_{1}(\mu
)B_{2}(\lambda )+\frac{x_{5}(\lambda -\mu )}{x_{1}(\lambda -\mu )}B_{2}(\mu
)B_{1}(\lambda )  \label{eq4.22}
\end{eqnarray}
Here we observe that in this approach the final action of $\tau (\lambda )$
on normally ordered states must be normal ordered. This implies in an
increasing use of commutation relations needed for the diagonalization of $%
\tau (\lambda )$. For example, the action of the operator $A_{1}(\lambda )$
on $\Psi _{2}(\lambda _{1},\lambda _{2})$ has the form 
\begin{eqnarray}
A_{1}(\lambda )\Psi _{2}(\lambda _{1},\lambda _{2}) &=&z(\lambda
_{10})z(\lambda _{20})x_{1}^{N}(\lambda )\ \Psi _{2}(\lambda _{1},\lambda
_{2})  \nonumber \\
&&-\frac{x_{5}(\lambda _{10})}{x_{2}(\lambda _{10})}z(\lambda
_{21})x_{1}^{N}(\lambda _{1})\ B_{1}(\lambda )B_{1}(\lambda _{2})\left|
0\right\rangle  \nonumber \\
&&-\frac{x_{5}(\lambda _{20})}{x_{2}(\lambda _{20})}\frac{z(\lambda _{12})}{%
\omega (\lambda _{12})}x_{1}^{N}(\lambda _{2})\ B_{1}(\lambda )B_{1}(\lambda
_{1})\left| 0\right\rangle  \nonumber \\
&&+\left( \frac{z(\lambda _{10})}{\omega (\lambda _{10})}\frac{x_{5}(\lambda
_{20})}{x_{2}(\lambda _{20})}\frac{1}{y(\lambda _{01})}+\frac{x_{7}(\lambda
_{10})}{x_{3}(\lambda _{10})}\frac{1}{y(\lambda _{12})}\right)  \nonumber \\
&&\times x_{1}^{N}(\lambda _{1})x_{1}^{N}(\lambda _{2})\ B_{2}(\lambda
)\left| 0\right\rangle  \label{eq4.23}
\end{eqnarray}
where $\lambda _{ab}=\lambda _{a}-\lambda _{b}$, $a\neq b=0,1,2$, with $%
\lambda _{0}=\lambda $. Here we have used the following identities satisfied
by the matrix elements of these $19$-vertex models: 
\begin{eqnarray}
\frac{z(\lambda _{ab})}{\omega (\lambda _{ab})}\frac{x_{5}(\lambda _{cb})}{%
x_{2}(\lambda _{cb})}-\varepsilon \frac{x_{6}(\lambda _{ab})}{x_{3}(\lambda
_{ab})}\frac{1}{y(\lambda _{ac})} &=&\frac{x_{5}(\lambda _{ab})}{%
x_{2}(\lambda _{ab})}\frac{x_{5}(\lambda _{ca})}{x_{2}(\lambda _{ca})}+\frac{%
z(\lambda _{ac})}{\omega (\lambda _{ac})}\frac{x_{5}(\lambda _{cb})}{%
x_{2}(\lambda _{cb})},  \nonumber \\
z(\lambda _{ab})\frac{x_{5}(\lambda _{cb})}{x_{2}(\lambda _{cb})}\frac{1}{%
y(\lambda _{ab})}+\frac{x_{1}(\lambda _{ab})}{x_{3}(\lambda _{ab})}\frac{1}{%
y(\lambda _{ac})} &=&z(\lambda _{ab})z(\lambda _{cb})\frac{1}{y(\lambda
_{ac})},  \nonumber \\
\omega (\lambda _{ab})\omega (\lambda _{ba}) &=&1,\qquad \qquad (a\neq b\neq
c)  \label{eq4.24}
\end{eqnarray}
Similarly, for the operator $A_{2}(\lambda )$ we have 
\begin{eqnarray}
&&A_{2}(\lambda )\Psi _{2}(\lambda _{1},\lambda _{2})=\quad \varepsilon ^{2}%
\frac{z(\lambda _{01})}{\omega (\lambda _{01})}\frac{z(\lambda _{02})}{%
\omega (\lambda _{02})}x_{2}^{N}(\lambda )\ \Psi _{2}(\lambda _{1},\lambda
_{2})  \nonumber \\
&&-\varepsilon ^{2}\frac{y_{5}(\lambda _{02})}{x_{2}(\lambda _{02})}%
z(\lambda _{21})x_{2}^{N}(\lambda _{2})\ B_{1}(\lambda )B_{1}(\lambda
_{1})\left| 0\right\rangle  \nonumber \\
&&-\varepsilon ^{2}\frac{y_{5}(\lambda _{01})}{x_{2}(\lambda _{01})}\frac{%
z(\lambda _{12})}{\omega (\lambda _{12})}x_{2}^{N}(\lambda _{1})\
B_{1}(\lambda )B_{1}(\lambda _{2})\left| 0\right\rangle  \nonumber \\
&&+z(\lambda _{21})\frac{1}{y(\lambda _{01})}x_{1}^{N}(\lambda _{1})\
B_{3}(\lambda )B_{1}(\lambda _{2})\left| 0\right\rangle  \nonumber \\
&&+\frac{z(\lambda _{12})}{\omega (\lambda _{12})}\frac{1}{y(\lambda _{02})}%
x_{1}^{N}(\lambda _{2})\ B_{3}(\lambda )B_{1}(\lambda _{1})\left|
0\right\rangle  \nonumber \\
&&+\varepsilon ^{2}\frac{y_{5}(\lambda _{01})}{x_{2}(\lambda _{01})}\left( 
\frac{y_{5}(\lambda _{21})}{x_{2}(\lambda _{21})}\frac{1}{y(\lambda _{01})}+%
\frac{z(\lambda _{01})}{\omega (\lambda _{01})}\frac{1}{y(\lambda _{02})}-%
\frac{y_{5}(\lambda _{01})}{x_{2}(\lambda _{01})}\frac{1}{y(\lambda _{12})}%
\right)  \nonumber \\
&&\times x_{1}^{N}(\lambda _{2})x_{2}^{N}(\lambda _{1})\ B_{2}(\lambda
)\left| 0\right\rangle  \nonumber \\
&&+\varepsilon ^{2}\frac{1}{y(\lambda _{01})}\left( z(\lambda _{01})\frac{%
y_{5}(\lambda _{02})}{x_{2}(\lambda _{02})}-\frac{y_{5}(\lambda _{01})}{%
x_{2}(\lambda _{01})}\frac{y_{5}(\lambda _{02})}{x_{2}(\lambda _{02})}%
\right) x_{1}^{N}(\lambda _{1})x_{2}^{N}(\lambda _{2})\ B_{2}(\lambda
)\left| 0\right\rangle  \label{eq4.25}
\end{eqnarray}
In this case we have used more two identities: 
\begin{eqnarray}
\frac{z(\lambda _{ab})}{\omega (\lambda _{ab})}\frac{1}{y(\lambda _{ac})}+%
\frac{y_{5}(\lambda _{cb})}{x_{2}(\lambda _{cb})}\frac{1}{y(\lambda _{ab})}
&=&\frac{y_{5}(\lambda _{ab})}{x_{2}(\lambda _{ab})}\frac{1}{y(\lambda _{bc})%
}+\frac{z(\lambda _{bc})}{\omega (\lambda _{bc})}\frac{1}{y(\lambda _{ac})} 
\nonumber \\
z(\lambda _{cb})\frac{y_{5}(\lambda _{ac})}{x_{2}(\lambda _{ac})}+\frac{%
y_{5}(\lambda _{ab})}{x_{2}(\lambda _{ab})}\frac{y_{5}(\lambda _{bc})}{%
x_{2}(\lambda _{bc})} &=&\frac{z(\lambda _{ab})y_{5}(\lambda _{ac})}{%
x_{2}(\lambda _{ac})}  \nonumber \\
a &\neq &b\neq c  \label{eq4.26}
\end{eqnarray}
Finally, for $A_{3}(\lambda )$ we get 
\begin{eqnarray}
&&\left. A_{3}(\lambda )\Psi _{2}(\lambda _{1},\lambda _{2})=\frac{%
x_{2}(\lambda _{01})}{x_{3}(\lambda _{01})}\frac{x_{2}(\lambda _{02})}{%
x_{3}(\lambda _{02})}x_{3}^{N}(\lambda )\ \Psi _{2}(\lambda _{1},\lambda
_{2})\right.  \nonumber \\
&&\left. -\varepsilon ^{2}\frac{z(\lambda _{12})}{\omega (\lambda _{12})}%
\frac{1}{y(\lambda _{01})}x_{2}^{N}(\lambda _{1})\ B_{3}(\lambda
)B_{1}(\lambda _{2})\left| 0\right\rangle -\varepsilon ^{2}z(\lambda _{21})%
\frac{1}{y(\lambda _{02})}x_{2}^{N}(\lambda _{2})\ B_{3}(\lambda
)B_{1}(\lambda _{1})\left| 0\right\rangle \right.  \nonumber \\
&&+\left( \frac{y_{7}(\lambda _{01})}{x_{3}(\lambda _{01})}\frac{1}{%
y(\lambda _{12})}-\frac{y_{5}(\lambda _{01})}{x_{3}(\lambda _{01})}\frac{1}{%
y(\lambda _{02})}\right) x_{2}^{N}(\lambda _{1})x_{2}^{N}(\lambda _{2})\
B_{2}(\lambda )\left| 0\right\rangle  \label{eq4.27}
\end{eqnarray}
Here we also have used the identities (\ref{eq4.24}) and (\ref{eq4.26}).

From these relations one can see that all unwanted terms of $\tau (\lambda
)\Psi _{2}(\lambda _{1},\lambda _{2})$ vanish. It means that $\Psi
_{2}(\lambda _{1},\lambda _{2})$ is an eigenstate of the transfer matrix $%
\tau (\lambda )$ with eigenvalue

\begin{equation}
\Lambda _{2}(\lambda ,\lambda _{1},\lambda _{2})=z(\lambda _{10})z(\lambda
_{20})x_{1}^{N}(\lambda )+\varepsilon ^{3}\frac{z(\lambda _{01})}{\omega
(\lambda _{01})}\frac{z(\lambda _{02})}{\omega (\lambda _{02})}%
x_{2}^{N}(\lambda )+\frac{x_{2}(\lambda _{01})}{x_{3}(\lambda _{01})}\frac{%
x_{2}(\lambda _{02})}{x_{3}(\lambda _{02})}x_{3}^{N}(\lambda )
\label{eq4.28}
\end{equation}
provided the rapidities $\lambda _{1}$ and $\lambda _{2}$ satisfy the 
{\small BA} equations 
\begin{equation}
\left( \frac{x_{1}(\lambda _{a})}{x_{2}(\lambda _{a})}\right)
^{N}=\varepsilon ^{3}\frac{z(\lambda _{ab})}{z(\lambda _{ba})}\omega
(\lambda _{ba})\ ,\quad a\neq b=1,2.  \label{eq4.29}
\end{equation}

\subsection{General Sector}

The generalization of the above results to sectors with more than two
particles proceeds through the factorization properties of the higher order
phase shifts discussed in the previous section. Therefore, at this point we
shall present the general result: In a generic sector $M=n$ , we have $n-1$
swap conditions 
\begin{equation}
\Psi _{n}(\lambda _{1},\cdots ,\lambda _{i-1},\lambda _{i+1},\lambda
_{i},\cdots ,\lambda _{n})=\omega (\lambda _{i}-\lambda _{i+1})\Psi
_{n}(\lambda _{1},\cdots ,\lambda _{i-1},\lambda _{i},\lambda _{i+1},\cdots
,\lambda _{n})  \label{eq4.30}
\end{equation}
which yield the $n-1$ operator-valued functions $\Gamma _{i}(\lambda
_{1},\cdots ,\lambda _{n})$ . The corresponding normal ordered state $\Psi
_{n}(\lambda _{1},\cdots ,\lambda _{n})$ can be written with aid of a
recurrence formula \cite{TA}: 
\begin{equation}
\Psi _{n}(\lambda _{1},...,\lambda _{n})=\Phi _{n}(\lambda _{1},...,\lambda
_{n})\left| 0\right\rangle  \label{eq4.31}
\end{equation}
where 
\begin{eqnarray}
&&\left. \Phi _{n}(\lambda _{1},...,\lambda _{n})=B_{1}(\lambda _{1})\Phi
_{n-1}(\lambda _{2},...,\lambda _{n})\right.  \nonumber \\
&&\left. -B_{2}(\lambda _{1})\sum_{j=2}^{n}\frac{1}{y(\lambda _{1}-\lambda
_{j})}\prod_{k=2,k\neq j}^{n}{\cal Z}(\lambda _{k}-\lambda _{j})\Phi
_{n-2}(\lambda _{2},...,\stackrel{\wedge }{\lambda }_{j},...,\lambda
_{n})A_{1}(\lambda _{j})\right.  \label{eq4.32}
\end{eqnarray}
with the initial condition $\Phi _{0}=1,\quad \Phi _{1}(\lambda
)=B_{1}(\lambda )$.

The scalar function ${\cal Z}(\lambda _{k}-\lambda _{j})$ is defined by 
\begin{equation}
{\cal Z}(\lambda _{k}-\lambda _{j})=\left\{ 
\begin{array}{c}
z(\lambda _{k}-\lambda _{j})\qquad \qquad \quad \quad {\rm if}\quad k>j \\ 
z(\lambda _{k}-\lambda _{j})\omega (\lambda _{j}-\lambda _{k})\quad \ {\rm if%
}\quad k<j
\end{array}
\right.  \label{eq4.33}
\end{equation}

The action of the operators $A_{i}(\lambda ),i=1,2,3$ on the operators $\Phi
_{n}$ have the following normal ordered form

\begin{eqnarray}
&&\left. A_{1}(\lambda )\Phi _{n}(\lambda _{1},...,\lambda
_{n})=\prod_{k=1}^{n}z(\lambda _{k}-\lambda )\Phi _{n}(\lambda
_{1},...,\lambda _{n})A_{1}(\lambda )\right.  \nonumber \\
&&\left. -B_{1}(\lambda )\sum_{j=1}^{n}\frac{x_{5}(\lambda _{j}-\lambda )}{%
x_{2}(\lambda _{j}-\lambda )}\prod_{k=1,k\neq j}^{n}{\cal Z}(\lambda
_{k}-\lambda _{j})\Phi _{n-1}(\lambda _{1},...,\stackrel{\wedge }{\lambda }%
_{j},...,\lambda _{n})A_{1}(\lambda _{j})\right.  \nonumber \\
&&\left. +B_{2}(\lambda )\sum_{j=2}^{n}\sum_{l=1}^{j-1}G_{jl}(\lambda
,\lambda _{l},\lambda _{j})\prod_{k=1,k\neq j,l}^{n}{\cal Z}(\lambda
_{k}-\lambda _{l}){\cal Z}(\lambda _{k}-\lambda _{j})\right.  \nonumber \\
&&\left. \times \Phi _{n-2}(\lambda _{1},...,\stackrel{\wedge }{\lambda }%
_{l},...,\stackrel{\wedge }{\lambda }_{j},...,\lambda _{n})A_{1}(\lambda
_{l})A_{1}(\lambda _{j})\right.  \label{eq4.34}
\end{eqnarray}
where $G_{jl}(\lambda ,\lambda _{l},\lambda _{j})$ are scalar functions
defined by 
\begin{equation}
G_{jl}(\lambda ,\lambda _{l},\lambda _{j})=\frac{x_{7}(\lambda _{l}-\lambda )%
}{x_{3}(\lambda _{l}-\lambda )}\frac{1}{y(\lambda _{l}-\lambda _{j})}+\frac{%
z(\lambda _{l}-\lambda )}{\omega (\lambda _{l}-\lambda )}\frac{x_{5}(\lambda
_{j}-\lambda )}{x_{2}(\lambda _{j}-\lambda )}\frac{1}{y(\lambda -\lambda
_{l})}  \label{eq4.37}
\end{equation}
For the action of $A_{3}(\lambda )$ we have a similar expression 
\begin{eqnarray}
&&\left. A_{3}(\lambda )\Phi _{n}(\lambda _{1},...,\lambda
_{n})=\prod_{k=1}^{n}\frac{x_{2}(\lambda -\lambda _{k})}{x_{3}(\lambda
-\lambda _{k})}\Phi _{n}(\lambda _{1},...,\lambda _{n})A_{3}(\lambda )\right.
\nonumber \\
&&\left. -\varepsilon ^{n}B_{3}(\lambda )\sum_{j=1}^{n}\frac{1}{y(\lambda
-\lambda _{j})}\prod_{k=1,k\neq j}^{n}{\cal Z}(\lambda _{j}-\lambda
_{k})\Phi _{n-1}(\lambda _{1},...,\stackrel{\wedge }{\lambda }%
_{j},...,\lambda _{n})A_{2}(\lambda _{j})\right.  \nonumber \\
&&\left. +B_{2}(\lambda )\sum_{j=2}^{n}\sum_{l=1}^{j-1}H_{jl}(\lambda
,\lambda _{l},\lambda _{j})\prod_{k=1,k\neq j,l}^{n}{\cal Z}(\lambda
_{j}-\lambda _{k}){\cal Z}(\lambda _{l}-\lambda _{k})\right.  \nonumber \\
&&\times \Phi _{n-2}(\lambda _{1},...,\stackrel{\wedge }{\lambda }_{l},...,%
\stackrel{\wedge }{\lambda }_{j},...,\lambda _{n})A_{2}(\lambda
_{l})A_{2}(\lambda _{j})  \label{eq4.36}
\end{eqnarray}
where the scalar functions $H_{jl}(\lambda ,\lambda _{l},\lambda _{j})$ are
given by 
\begin{equation}
H_{jl}(\lambda ,\lambda _{l},\lambda _{j})=\frac{y_{7}(\lambda -\lambda _{l})%
}{x_{3}(\lambda -\lambda _{l})}\frac{1}{y(\lambda _{l}-\lambda _{j})}-\frac{%
y_{5}(\lambda -\lambda _{l})}{x_{3}(\lambda -\lambda _{l})}\frac{1}{%
y(\lambda -\lambda j)}  \label{eq4.50}
\end{equation}
The action of the operator $A_{2}(\lambda )$ is more cumbersome 
\begin{eqnarray}
&&\left. A_{2}(\lambda )\Phi _{n}(\lambda _{1},...,\lambda _{n})=\varepsilon
^{n}\prod_{k=1}^{n}\frac{z(\lambda -\lambda _{k})}{\omega (\lambda -\lambda
_{k})}\Phi _{n}(\lambda _{1},...,\lambda _{n})A_{2}(\lambda )\right. 
\nonumber \\
&&\left. -\varepsilon ^{n}B_{1}(\lambda )\sum_{j=1}^{n}\frac{y_{5}(\lambda
-\lambda _{j})}{x_{2}(\lambda -\lambda _{j})}\prod_{k=1,k\neq j}^{n}{\cal Z}%
(\lambda _{j}-\lambda _{k})\Phi _{n-1}(\lambda _{1},...,\stackrel{\wedge }{%
\lambda }_{j},...,\lambda _{n})A_{2}(\lambda _{j})\right.  \nonumber \\
&&\left. +B_{3}(\lambda )\sum_{j=1}^{n}\frac{1}{y(\lambda -\lambda _{j})}%
\prod_{k=1,k\neq j}^{n}{\cal Z}(\lambda _{k}-\lambda _{j})\Phi
_{n-1}(\lambda _{1},...,\stackrel{\wedge }{\lambda }_{j},...,\lambda
_{n})A_{1}(\lambda _{j})\right.  \nonumber \\
&&+\varepsilon ^{n}B_{2}(\lambda )\left\{
\sum_{j=2}^{n}\sum_{l=1}^{j-1}Y_{jl}(\lambda ,\lambda _{l},\lambda
_{j})\prod_{k=1,k\neq j,l}^{n}{\cal Z}(\lambda _{k}-\lambda _{l}){\cal Z}%
(\lambda _{j}-\lambda _{k})\right. \times  \nonumber \\
&&\left. \Phi _{n-2}(\lambda _{1},...,\stackrel{\wedge }{\lambda }_{l},...,%
\stackrel{\wedge }{\lambda }_{j},...,\lambda _{n})A_{1}(\lambda
_{l})A_{2}(\lambda _{j})\right. +  \nonumber \\
&&\left. \sum_{j=2}^{n}\sum_{l=1}^{j-1}F_{jl}(\lambda ,\lambda _{l},\lambda
_{j})\prod_{k=1,k\neq j,l}^{n}{\cal Z}(\lambda _{l}-\lambda _{k}){\cal Z}%
(\lambda _{k}-\lambda _{j})\right. \times  \nonumber \\
&&\left. \Phi _{n-2}(\lambda _{1},...,\stackrel{\wedge }{\lambda }_{l},...,%
\stackrel{\wedge }{\lambda }_{j},...,\lambda _{n})A_{1}(\lambda
_{j})A_{2}(\lambda _{l})\right\}  \label{eq4.35}
\end{eqnarray}
where we have more two scalar functions

\begin{eqnarray}
F_{jl}(\lambda ,\lambda _{l},\lambda _{j}) &=&\frac{y_{5}(\lambda -\lambda
_{l})}{x_{2}(\lambda -\lambda _{l})}\left\{ \frac{y_{5}(\lambda _{l}-\lambda
_{j})}{x_{2}(\lambda _{l}-\lambda _{j})}\frac{1}{y(\lambda -\lambda _{l})}+%
\frac{z(\lambda -\lambda _{l})}{\omega (\lambda -\lambda _{l})}\frac{1}{%
y(\lambda -\lambda _{j})}\right.  \nonumber \\
&&\left. -\frac{y_{5}(\lambda -\lambda _{l})}{x_{2}(\lambda -\lambda _{l})}%
\frac{1}{y(\lambda _{l}-\lambda _{j})}\right\}  \label{eq4.38}
\end{eqnarray}
\begin{equation}
Y_{jl}(\lambda ,\lambda _{l},\lambda _{j})=\frac{1}{y(\lambda -\lambda _{l})}%
\left\{ z(\lambda -\lambda _{l})\frac{y_{5}(\lambda -\lambda _{j})}{%
x_{2}(\lambda -\lambda _{j})}-\frac{y_{5}(\lambda -\lambda _{l})}{%
x_{2}(\lambda -\lambda _{l})}\frac{y_{5}(\lambda _{l}-\lambda _{j})}{%
x_{2}(\lambda _{l}-\lambda _{j})}\right\}  \label{eq4.39}
\end{equation}

From these relations immediately follows that $\Psi _{M}(\lambda
_{1},...,\lambda _{M})$ are the eigenstates of $\tau (\lambda )$ with
eigenvalues 
\begin{equation}
\Lambda _{M}=x_{1}(\lambda )^{N}\prod_{a=1}^{M}z(\lambda _{a}-\lambda
)+\varepsilon ^{M+1}x_{2}(\lambda )^{N}\prod_{a=1}^{M}\frac{z(\lambda
-\lambda _{a})}{\omega (\lambda -\lambda _{a})}+x_{3}(\lambda
)^{N}\prod_{a=1}^{M}\frac{x_{2}(\lambda -\lambda _{a})}{x_{3}(\lambda
-\lambda _{a})}  \label{eq4.51}
\end{equation}
provided their rapidities $\lambda _{i},i=1,...,M$ \ satisfy the {\small BA}
equations 
\begin{equation}
\left( \frac{x_{1}(\lambda _{a})}{x_{2}(\lambda _{a})}\right)
^{N}=\varepsilon ^{M+1}\prod_{b\neq a=1}^{M}\frac{z(\lambda _{a}-\lambda
_{b})}{z(\lambda _{b}-\lambda _{a})}\omega (\lambda _{b}-\lambda _{a}),\quad
a=1,2,...,M  \label{eq4.52}
\end{equation}
To conclude this section we remark that equations (\ref{eq4.51}) and (\ref
{eq4.52}) reproduce the known results in the literature: Using (\ref{RIK})
we reproduce the {\small BA} solution of the {\small IK}\ model \cite{TA}
and using (\ref{RZF}) they are the {\small BA} solution for the {\small ZF}
model derived by a fusion procedure in \cite{KRS}, \cite{BT} and \cite{KR} .
Specifically, in the case of the rational solution and for $\varepsilon =-1 $
we obtain previous results derived by the Analytical \cite{Ku2} and \
Algebraic \cite{MR} {\small BA} approach for the rational $osp(1|2)$ vertex
model. Furthermore, by using the equation (\ref{eq4.7}) we recover the
expressions derived in the previous section via the Coordinate {\small BA}.

\section{Conclusion}

In the first part of this paper we have applied the Coordinate {\small BA}
to find the spectra of Hamiltonians associated to three $19$-vertex models,
including a graded model. This procedure was carried out for periodic
boundary conditions.

We believe that the method here presented could also be applied for
Hamiltonians associated with higher states vertex-models. For instance, in
the quantum spin chain $s=3/2$ {\small XXZ} model we have four states: $%
\left| k[\frac{3}{2}]\right\rangle $, $\left| k[\frac{1}{2}]\right\rangle $, 
$\left| k[\frac{-1}{2}]\right\rangle $ and $\left| k[\frac{-3}{2}%
]\right\rangle .$ It means that the state $\left| k(1/2)\right\rangle $ can
be parametrized by plane wave and the states $\left| k[\frac{-1}{2}%
]\right\rangle $ and $\left| k[\frac{-3}{2}]\right\rangle $ as two and three
states $\left| k[\frac{1}{2}]\right\rangle $ at the same site ,
respectively, multiplied by some weight functions.

These weight functions are responsible by the factorized form of the phase
shift of two particle (\ref{eq3.47}). In the {\small ZF} model we do not a
factored form for the two-pseudoparticle phase shift because its weight
function (\ref{eq3.16}) is a constant. It means that the state $\left|
k[-]\right\rangle $ behaves exactly as two states $\left| k[0]\right\rangle $
at the same site. This is in agreement with the fact that the {\small ZF}
model can be constructed by a fusion procedure of two six-vertex model.

In the second part of this paper we have applied the Algebraic {\small BA}
to find the spectra of the transfer matrices of these three state vertex
models. The method here presented was developed by Tarasov \cite{TA} and
generalized by Martins \cite{MA, MR}. It is general enough to include the 
{\small ZF} model as well as the graded $osp(1|2)$ model.

There are several issues left for future works. A natural extension of this
work is to consider these Bethe Ans\"{a}tze with open boundary conditions
via reflection matrices. The transfer matrix of the {\small ZF} model with
the most general diagonal reflection matrix has been diagonalized by
Mezincescu {\it at al} \cite{Mez} by generalizing the fusion approach used
to solve the corresponding model with periodic boundaries. {\small BA}
equations for both the {\small ZF} model and {\small IK} model with open
boundaries were derived by Yung and Batchelor in ref.\cite{BAT2}.
Nevertheless, basead on the Tarasov-Martins approach, the Algebraic {\small %
BA} of the {\small IK} model with a diagonal $K$-matrix was recently derived
by Fan \cite{FAN}.

{\bf Acknowledgment:} The author have profited from discussions with R. Z.
Bariev and F.C. Alcaraz. This work was supported in part by Funda\c{c}\~{a}o
de Amparo \`{a} Pesquisa do Estado de S\~{a}o Paulo-{\small FAPESP}-Brasil
and Conselho Nacional de Desenvolvimento-{\small CNPq}-Brasil.

\newpage{}

\begin{center}
{\LARGE LIST OF CHANGES}
\end{center}

\vspace{2cm}
{}

Reference: A/99391/PAP

Title: Bethe Ansatze for nineteen-vertex models

Author: A. Lima-Santos

\vspace{0.5cm}

1. page 27, before the section 5 (Conclusion) we replace all text\ (lines
1,2 and 3): In particular, ...., via the Coordinate BA. by the following
text:

\vspace{0.3cm}

{\bf To conclude this section we remark that equations (4.50) and (4.51)
reproduce the known results in the literature: Using (2.8) we reproduce the
BA solution of the IK\ model [16] and using (2.7) they are the BA solution
for the ZF model derived by a fusion procedure in [13], [14] and [15] .
Specifically, in the case of the rational solution and for }$\varepsilon =-1$%
{\bf \ we obtain previous results derived by the Analytical [20] and \
Algebraic [23] BA approach for the rational }$osp(1|2)${\bf \ vertex model.
Furthermore, by using the equation (4.7) we recover the expressions derived
in the previous section via the Coordinate BA.}

\end{document}